\documentclass[]{RAA}

\usepackage{booktabs}
\usepackage{graphicx}
\usepackage{shorttoc}
\usepackage{natbib} 
\usepackage{lscape}
\usepackage{amsmath}
\usepackage{textcomp}
\usepackage{longtable}
\usepackage[colorlinks,linkcolor=black,anchorcolor=black,citecolor=blue]{hyperref}
\input{epsf.sty}
\input{psfig.sty} 
\begin{document}

\title{The First Data Release (DR1) of the LAMOST general survey}

\volnopage{Vol.0 (2010) No.0, 000--000}

\setcounter{page}{1}        
 \author{A.-L. Luo \inst{1*} \and Y.-H. Zhao \inst{1} \and G. Zhao \inst{1} \and L.-C. Deng \inst{1} \and X.-W. Liu \inst{5} \and Y.-P. Jing\inst{4}  \and G. Wang \inst{1} \and H.-T Zhang \inst{1} \and J.-R. Shi \inst{1}  \and X.-Q. Cui \inst{2}
% ...
\\  
  \institute{National Astronomical Observatories, Chinese Academy of Sciences, Beijing 100012, China \\
                   \ \  \ \  { \it * lal@nao.cas.cn  }
      \and   Nanjing Institute of Astronomical Optics and Technology, National Astronomical Observatories,  
                   Chinese Academy of Sciences, Nanjing 210042, China\\
     \and
            University of Science and Technology of China, Hefei 230026, China\\
      \and
           Shanghai Astronomical Observatory, Chinese Academy of Sciences, Shanghai 200030, China \\
      \and
           Department of Astronomy, Peking University, Beijing, 100871,China\\
      \and
           Department of Astronomy, Beijing Normal University, Beijing, 100875,China\\
     \and
           Department of Physics, Applied Physics and Astronomy, Rensselaer Polytechnic Institute, Troy, NY 12180, USA\\
     \and
           Department of Physics and Astronomy, Earlham College, Richmond, IN 47374, USA\\
      \and
           Universit\"{a}t Heidelberg, Zentrum f\"{u}r Astronomie, Landessternwarte, K\"{o}nigstuhl 12, 69117 Heidelberg, Germany\\
     \and 
           Yunnan Observatory, Chinese Academy of Sciences, Kunming, 650216, China\\
     \and 
           Purple Mountain Observatory, Chinese Academy of Sciences, Nanjing 210008, China \\
   }
}
\date{Received~~2014 month day; accepted~~2014~~month day}

\abstract{  The Large sky Area Multi-Object Spectroscopic Telescope (LAMOST) General Survey is a spectroscopic survey that will eventually cover approximately half of the celestial sphere and collect 10 million spectra of stars, galaxies and QSOs. Objects both in the pilot survey and the first year general survey are included in the LAMOST First Data Release (DR1).  The pilot survey started in October 2011 and ended in June 2012, and the data have been released to the public as the LAMOST Pilot Data Release in August 2012. The general survey started in September 2012, and completed its first year of operation in June 2013. The LAMOST DR1 includes a total of 1202 plates containing 2,955,336 spectra, of which 1,790,879 spectra have observed signal-to-noise ($S/N) \geq $10. All data with $S/N\geq $2 are formally released as LAMOST DR1 under the LAMOST data policy. This data release contains a total of 2,204,696 spectra, of which 1,944,329 are stellar spectra, 12,082 are galaxy spectra and 5,017 are quasars. The DR1 includes not only spectra, but also three stellar catalogues with measured parameters: AFGK-type stars with high quality spectra (1,061,918 entries), A-type stars (100,073 entries), and M stars (121,522 entries).  This paper introduces the survey design, the observational and instrumental limitations, data reduction and analysis, and some caveats. Description of the FITS structure of spectral files and parameter catalogues is also provided.
}
   \titlerunning{Data release one (DR1) of LAMOST general survey }
   \maketitle
\section{Introduction}
\label{sect:intro}

The Large sky Area Multi-Object Spectroscopic Telescope (LAMOST) is a 4.0m telescope located at the Xinglong Observatory northeast of Beijing, China. The telescope is characterized by both a large field of view and large aperture. A total of 4000 fibers mounted on its focal plane make the LAMOST telescope a powerful tool with high spectral acquisition rate \citep{2012RAA....12..735D}. With the capability of its large field of view and immense multiplexing, the LAMOST telescope is dedicated to a spectral survey of celestial objects over the entire available northern sky. After first light in 2008, commissioning observations were conducted for two years. In spite of the ongoing characterization of the LAMOST system, observation limitations (e.g., dome seeing, fiber positioning accuracy), and data processing software, the commissioning data have contributed some preliminary results on quasars \citep{2010RAA....10..612H,2010RAA....10..737W,2010RAA....10..745W}, planetary nebulae \citep{2010RAA....10..599Y}, and nearby galaxies \citep{2011RAA....11.1093Z}. 

In order to inspect the real instrumental performance and assess the feasibility of the science goals, the science working groups of LAMOST came up with the Pilot Survey, which was launched on 24 October 2011. In the Pilot Survey, both Galactic and extragalactic surveys were conducted -- these were named LEGUE (LAMOST Experiment for Galactic Understanding and Exploration) and LEGAS (LAMOST Extra GAlactic Survey), respectively \citep{2012RAA....12..723Z}. LEGUE is made up of three parts, which focus on the Galactic anti-center, the disk at longitudes away from the Galactic anti-center, and the halo. Some scientific results from the Pilot Survey include catalogs of white dwarf-main sequence binaries \citep{2013AJ....146...82R}, DA white dwarfs \citep{2013AJ....146...34Z,2013AJ....145..169Z}, and M dwarfs \citep{2014AJ....147...33Y}, and identification of velocity substructure in the Galactic disk \citep{2013ApJ...777L...5C}. Meanwhile, the LEGAS extragalactic portion of the survey studies galaxy redshifts, stellar populations of galaxies, and QSOs. Scientific results obtained include the identification of QSOs in the background of M31/M33 \citep{2013AJ....145..159H}, a study of the radio-loud quasar LAMOST J1131+3114 \citep{2014A&A...564A..89S}, and a search for double-peaked, narrow emission-line galaxies \citep{2014RAA....14.1234S}. The LAMOST Pilot Survey ended in June 2012, and included nine full moon cycles covering all good observing seasons at the site. The Pilot Survey data release includes 717,660 spectra, in which there are 648,820 stars, 2,723 galaxies, and 621 quasars, and a stellar parameter catalogue containing entries for 373,481 stars \citep{2012RAA....12.1243L}. 

From the experience of the Pilot Survey, the LAMOST science committee fully analyzed the observations, assessing the magnitude limit, characteristics of the telescope, and data success rate. A five-year General Survey of LAMOST was designed, which would mainly focus on the Galactic survey of stars, yet still include a significant number of extra-galactic sources.

As the only resolved classical disk galaxy, the Milky Way serves as a laboratory to study the formation and evolution of galaxies. However, there have been no large-scale Galactic spectroscopic surveys with a uniform sampling over all stellar populations, due to the time consuming nature of observing a large fraction of its $\sim$20 billion members spread over the entire sky. Thus our knowledge of the detailed structure of our Galaxy lags behind the understanding of the large-scale universe. Several notable achievements have been made by modern large-scale surveys such as Sloan Digital Sky Survey (SDSS); however, because SDSS is mainly focused on extragalactic studies, its spectroscopic observations have sampled only a small fraction of the Milky Way's stellar components. The powerful spectral acquisition rate of LAMOST enables a homogeneous and (statistically) complete Galactic survey.

The design of the survey will be shown in Section~2, including fields and sample selection. The observation conditions and limitations due the instruments will be discussed in Section~3. The LAMOST data reduction, spectral analysis, and stellar parameter pipelines will be introduced in Section~4. The data products of DR1 will be introduced in Section~5, as well as caveats. Finally, the data access and use are introduced in Section~6.

\section{design of the survey}

\subsection{LEGUE (LAMOST Experiment for Galactic Understanding and Exploration)}

As a major component of the LAMOST project, the LEGUE spectroscopic survey of the Galactic halo will observe $\sim$5.8~million objects with $r<16.8$~mag at $|b|>30^{\circ} $ located mostly in the sky area with SDSS photometric coverage \citep{2012RAA....12..781Y,2012RAA....12..735D,2012RAA....12..792Z}. Considering time consuming for faint objects which should be observed in dark nights, two medium-bright (M) plates stripes with high priority are proposed.  One is the area at Dec. $= 30^{\circ}$ in Northern Galactic cap, and the other is at Dec. $ = 5^{\circ}$ in Southern Galactic cap.

Another unique component of LEGUE is the LAMOST Spectroscopic Survey of the Galactic Anti--center (LSS-GAC), which aims to survey a significant volume over a continuous sky area of $\sim$ 3,400 deg$^2$, covering Galactic longitudes 150$^{\circ}\leq \ell \leq$210$^{\circ}$ and latitudes $|b|\leq30^{\circ}$ \citep{2014arXiv1412.6628Y}. The input catalog is selected from photometric catalogs of the Xuyi Schmidt Telescope Photometric Survey of the Galactic anti-center (XSTPS-GAC) \citep{2014IAUS..298..310L}, with most targets selected with uniform probability. 
The catalog will include a statistically complete sample of about 3~million stars with magnitudes ranging from $14.0 < r 17.8$ (and as deep as 18.5 mag for some fields), with rare objects of extreme colors targeted with high priorities. A detailed description of LSS--GAC is presented by \citealt{2014arXiv1412.6628Y}.

For the Galactic disk survey, eight plates are chosen near the Galactic plane ($|b| \leq20^{\circ}$)and nearly uniformly distributed along the Galactic longitude, as well as in the region 0$^{\circ} < \alpha < 67^{\circ}$ and 42$^{\circ} < \delta < 59^{\circ}$ \citep{2012RAA....12..805C}. This survey concentrates on open clusters (OCs), and is expected to obtain about 8.9~million stars down to a limiting magnitude of $ r \sim 16$~mag. The highest observational priority is given to known or potential members of OCs, and a total of 194 OCs have high priority. The remainder of the disk targets in the 11.3 $<$ I$_{mag}$ $<$ 16.3 range were uniformly selected over I and B$-$I color-magnitude space. Very high proper motion objects (7mas/yr) are removed from the target catalogs. Positions, proper motions, and magnitudes are taken from the PPMXL \citep{2010AJ....139.2440R} catalog.

\subsection{LEGAS (LAMOST ExtraGAlactic Survey)}

The extra-galactic survey (LEGAS) consists of two main parts -- a galaxy survey and a QSO survey. 
For galaxies, it is split into two regions (again, to avoid the Galactic plane). One is in the North Galactic Cap region, and shares the same footprint as the SDSS legacy ellipse. This survey mainly aims to observe those SDSS main sample galaxies (with $r<17.75$) that were not spectroscopically observed by SDSS due to fiber collisions \citep{2002AJ....123..485S}. Such ``missed'' SDSS galaxies are named as the complementary galaxy sample in the LAMOST survey. There are 68,722 such targets, which are given highest priority for fiber assignment on plates when possible. Because the number density of galaxies is very low, the remaining fibers on LEGAS plates are filled by LEGUE halo survey stars. 
%and assigned fibers with higher priority considrring their lower number density. 
The other region of the galaxy survey in LAMOST is the South Galactic Cap region (SGC). The LEGAS team has a long-term strategic goal of a galaxy survey in the SGC footprint of LAMOST ($b<-30^{\circ}$ and $\delta > -10^{\circ}$), which aims to take spectra of all galaxies with $r<18$, and a sample of blue galaxies down to $r<18.8$. Since this desired galaxy survey in the SGC requires much deeper observations than the main LAMOST stellar sample (currently, mostly limited to $r<16.8$), only a few dark nights per observing season are available for these galaxies. Therefore, the galaxy survey in SGC now focuses on a strip, ranging from $45^{\circ}< \alpha< 60^{\circ}$ and $0.5^{\circ}< \delta < 9.5^{\circ}$. Besides these two well-defined samples of galaxies, some bright-infrared galaxies are selected as extra observational targets which are selected from infrared surveys such as IRAS, WISE, and HERSCHEL.

The efficiency of identifying quasars with redshifts between $2.2<z<3$ is low in SDSS \citep{2010yCat.7260....0S}, because quasars with such redshifts usually have similar optical colors as stars and are thus mostly ignored by the SDSS quasar candidate selection algorithm. As permitted by the observational conditions and instrument standards of the LAMOST telescope, LAMOST quasar survey aims to find as many quasars as possible. We selected quasar targets from optical data within the Sloan photometric footprint using data mining methods such as support vector machines (SVM) \citep{2012MNRAS.425.2599P, 2003PASP..115.1006Z}, distribution of quasars and stars in different color spaces including $Y-K$ vs. $g-z$ \citep{2012AJ....144...49W}) and $z-W1$ vs. $g-z$ \citep{2010MNRAS.406.1583W}, and the extreme-deconvolution method \citep{2011ApJ...729..141B}. The targets selected by multiple methods are given much higher priority in observation. The core sample is contributed from the above methods using survey data from SDSS, UKIDSS, WISE. The bonus sample comes from other survey data, such as X-ray bands from ROSAT, XMM, and Chandra, and radio bands from FIRST and NVSS. It is noted that the QSO targets are not limited to the LEGAS plates, and are given high priority in the LEGUE plates as well.

\subsection{The special fields}
\subsubsection{M31-M33}

Targets in the M31-M33 region include known and candidate planetary nebulae (PNe), HII regions, super--giants, and globular clusters in M31 and M33, plus known and candidate background quasars and foreground Galactic stars selected from the XSTPS-GAC photometric catalogs. These stars are selected, assigned and observed in the same way as for the LSS-GAC main survey. The selection criteria of various types of targets and candidates of interest are presented elsewhere (e.g., \citealt{2010RAA....10..599Y} for PNe; \citealt{2010RAA....10..612H} for quasars). The field central stars for the M31-M33 survey are selected similarly to the LSS-GAC main survey in the (RA, Dec) plane. Two groups of field central stars are chosen to cover the M31-M33 area (0$^{\circ}\leq$~RA~$\leq30^{\circ}$, 25$^{\circ} \leq$~Dec~$\leq 50^{\circ}$) independently. Each group contains 41 central stars.

\subsubsection{The LAMOST Complete Spectroscopic Survey of Pointing Area at Southern Galactic Cap} 
  
A LAMOST Key Project named the LAMOST Complete Spectroscopic Survey of Pointing Area (LCSSPA) at SGC is being conducted \citep{Yang..MNRAS..accepted}, which is designed to have repeated spectroscopic observations of all sources (Galactic and extra-galactic) in two 20~deg$^2$ fields in the SGC. The central coordinates of the fields are $(\alpha, \delta) = (37.88150939^\circ, 3.43934500^\circ)$ and $(21.525988792^\circ, -2.200949833^\circ)$, respectively. Targets in these field mainly consist of stars, galaxies, $u$-band variables, and QSOs. The stars and galaxies are selected from SDSS Data Release Nine (DR9) using $r$-band {\it psf} magnitudes and {\it petro} mags between $14.0 < r < 18.1$ mag, respectively. The $u$-band variable sources are selected as objects for which the difference in magnitudes between the SDSS and South Galactic Cap $u$-band Sky Survey (SCUSS) are larger than 0.2 mag; these are a magnitude-limited sample down to $u=19.0$ mag. The QSOs are selected by the same methods discussed above. The highest priority is assigned to $u$-band variables and QSOs, with galaxies are the second priority level, and stars having the lowest priority.

\subsubsection{Kepler fields}
Additionally, a total of 14 Kepler fields with a fixed field of view (FoV) of 105 square degrees in the constellation Lyra and Cygnus are also selected according to the prioritized list provided by LK-project \citep{De..Unknown..inpre}. The prioritized lists are designed mainly based on the type of targets and brightness. According to the object types, these targets are divided into four priority levels: standard targets ($\sim$250, MK secondary standard stars), KASC targets ($\sim$7,000, selected by the Kepler Asteroseismic Science Consortium, or KASC), planet targets ($\sim$150,000, selected by the Kepler planet search group \citealt{2010ApJ...713L.109B}), and extra targets ($\sim$1,000,000 other targets in the Kepler FoV from the KIC 10). For these, the higher observation priority is given to the fainter objects.

\subsection{Plate design}
\subsubsection{Plate definition and their magnitude }
The LAMOST plates are divided into four modes, very bright (VB), bright (B), medium-bright (M), and faint (F) according the magnitude range of stars to be observed. In DR1, the VB mode contains 2693 plates, and is used to observe stars brighter than $r = 14$~mag over the entire sky observable by LAMOST (-10$^{\circ}\leq$~Dec~$\leq$60$^{\circ}$ ) in order to make use of bright nights or nights of unfavorable conditions (e.g., large seeing or low atmospheric transparency). In the XSTPS-GAC footprint, stars are selected as potential targets from the XSTPS-GAC and 2MASS catalogs with magnitudes $r\leq14.0$~mag and  $9.0\leq~J~\leq12.5$, respectively. Stars outside of the XSTPS-GAC sky area are selected as potential targets from the PPMXL \citep{2010AJ....139.2440R} catalog with magnitude $10.0\leq~b1~\leq15.0$, $10.0 \leq~b2~\leq15.0$, $9.0\leq~r1~\leq14.0$, $9.0\leq~r2~\leq14.0$, or $8.5\leq~i~\leq13.5$, and from the 2MASS catalog with magnitude $9.0\leq~J~\leq12.5$. The ``B'' (bright) observing mode contains 751 plates in DR1 that observed stars with $r$ magnitudes between between 14 and 16.8. There are 692 and 655 plates assigned to the M ($16.8\leq r \leq17.8$), and F ($r>17.8$) designations, respectively.

\subsubsection{Tiling method \citep{2012RAA....12.1197C}}
Merging all of the science goals described above, the science committee of LAMOST designed the survey strategies and input catalogues. Based on these, the LAMOST team developed a software named SSS (Sky Strategy System) to produce observational plates that most efficiently use the multi-fiber abilities of LAMOST. The maximum density algorithm, density gradient algorithm, and the mean--shift algorithm are used for most sky areas, with the exception of the Galactic Anti--Center Survey, which uses separate target selection criteria \citep{2014arXiv1412.6628Y}.

Given a sky area, the maximum density method can find the tile (or plate) with the maximum object count by making a uniform tiling. By traversing all the tiles covering the sky area, a tile with the maximum object count will be found and then a new observational plate will be built at that position. The only factor that affects the speed and accuracy of this algorithm is the target distribution in the sky area. The denser the covering is, the higher the accuracy of finding the maximum tile becomes and the more efficient is the method, and vice--versa. 
The density gradient method is essentially iterative. Based on a uniform covering, this method makes an estimate of the object distribution and moves the tile in the direction that is expected to have a higher object density.
The basic idea of the mean-shift algorithm is also iterative. As long as the objects are distributed unevenly in the sky areas, a density gradient exists between different positions. Vectors are used to describe the position of targets in the tile, and the sum of vectors is set along the direction of the density gradient. By iteratively moving the field center along the sum of vectors and calculating the new vector summation, the recent peak density location is obtained. In order to maintain the goal of uniformity of the remaining area and to avoid neglecting some of the peak positions in areas where density varies intricately, the algorithm will try to find a position with a global higher density and make a quick coverage on the whole available sky area before iteration \citep{2007AcASn..48..500R}.

For each plate, one bright star (the so called central star, brighter than 8~mag) must be placed in the plate center to feed a Shack-Hartmann detector that drives the active optics system, and four stars (brighter than 17~mag) must be placed on each of four guiding CCD cameras. The bright stars located in the center were selected from Hipparcos catalog \citep{1997A&A...323L..49P}, with binaries excluded, and at the time of observation, the focal plane is rotated a bit to confirm that the proper guiding stars are inside the FoV of the four guiding CCD cameras. Guide stars were selected from the GSC2.3 catalog \citep{2008AJ....136..735L}. Central stars will be excluded if there are no corresponding guide stars. Since in some areas (e.g., at low Galactic latitudes) there are many more available central stars than needed to fully cover that part of sky, a step by step search was applied to reject redundant bright stars and find the tiling that maximizes sky coverage of that area \citep{2012SPIE.8448E..2AY}.

\subsubsection{Fiber allocation scheme}
Once the central position of a plate is decided by the tilling method, each object in the input catalog will be assigned a priority according to its scientific requirements. Five flux standards and 20 sky fibers are reserved in each spectrograph for data reduction; therefore, these five flux standards are allocated with the highest priority. All of the scientific targets are assigned in order of priority; however, if more than two targets have the same priority in the same fiber cell, the closest target will be chosen to avoid collision with other fibers. On the other hand, if a target can be reached by different fibers at the same time, then it will be assigned to the fiber within its fiber cell that contains the minimum number of possible targets \citep{2014IAUS..298..452Y}. A total of 20 sky fibers will eventually be assigned from fibers for which no targets were assigned. If the number of sky fibers allocated to a given spectrograph is less than 20, the lowest priority fibers will be reassigned to the sky until there are 20 sky fibers.

\section{limitations on the observations and performance of the telescope}

\subsection{Seeing }

One of the significant factors affecting the observations is seeing, which consists of the native site seeing as well as dome seeing. The seeing depends on the season, and is worse in the windy winter, but with a huge diversity. The peak in the distribution of the nightly site seeing is at around 3~arcsec, and 85\% of the seeing measurements are better than 4~arcsec \citep{2012RAA....12..772Y}. The non--classical dome of the LAMOST telescope seemingly affects the seeing measured directly at the focal plane, with the measurements showing seeing of about 3.5~arcsec on average. Compared to the site seeing, it is dome seeing that contributes the most in limitating the system capabilities (especially affecting the limiting magnitude). Considering the fiber diameter of 3~arcsec, an average increase of 1~arcsec in seeing (at the focal plane) will lead to a 0.4-magnitude decrease in flux collected by each fiber. 

\subsection{Fiber positioning and use efficiency}
The precision of fiber positioning is an important factor affecting the efficiency of the survey. Through repeated observations and calculations, the accuracy of fiber positioning has been determined to be less than 1.5~arcsec on average \citep{2014SPIE.9149E..1NC}, which has less effect on the fiber magnitude than the seeing.
Another limitation is provided by the fixed fiber density of around 200 per deg$^2$. This limits sampling in both dense and sparse fields in different sky areas. For dense fields, objects are too crowded to resolve (especially near objects such as stellar clusters). In contrast, the fiber density results in a considerable waste of fibers (which are allocated to sky) for sparse fields in high Galactic latitude areas.

\subsection{The performance of the instruments}

The performance of the instruments plays the same important role in the quantity and quality of spectra observed as the procedure of data reductions. We briefly describe it as follows.\\\\

\begin{figure}[h]
\centering
\includegraphics[width=70mm]{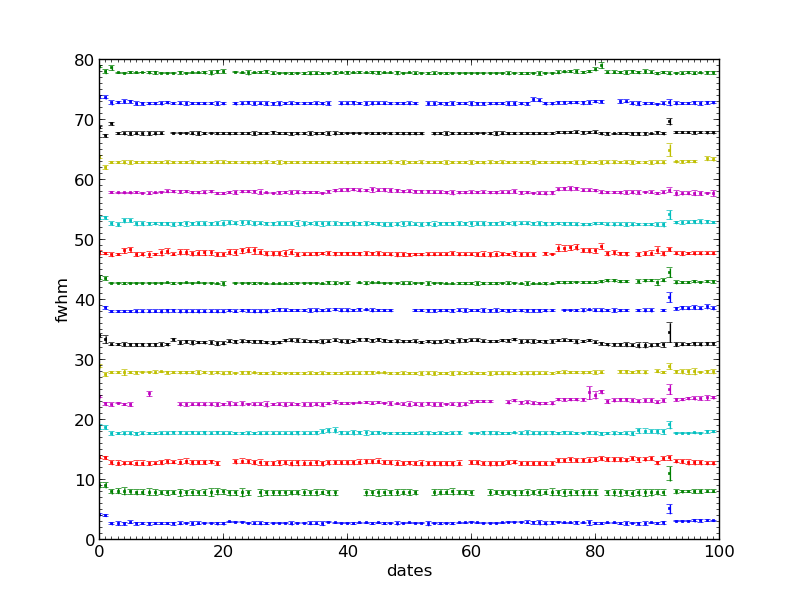}
\includegraphics[width=70mm]{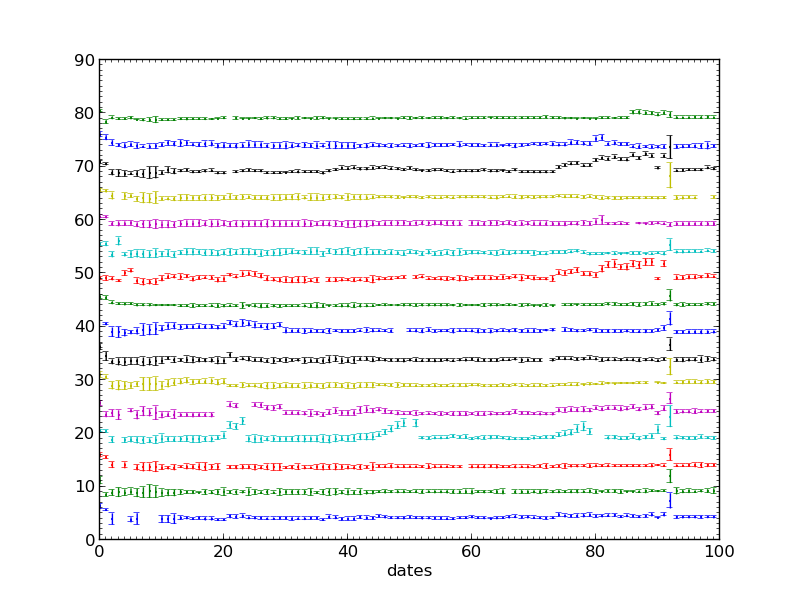}
\includegraphics[width=70mm]{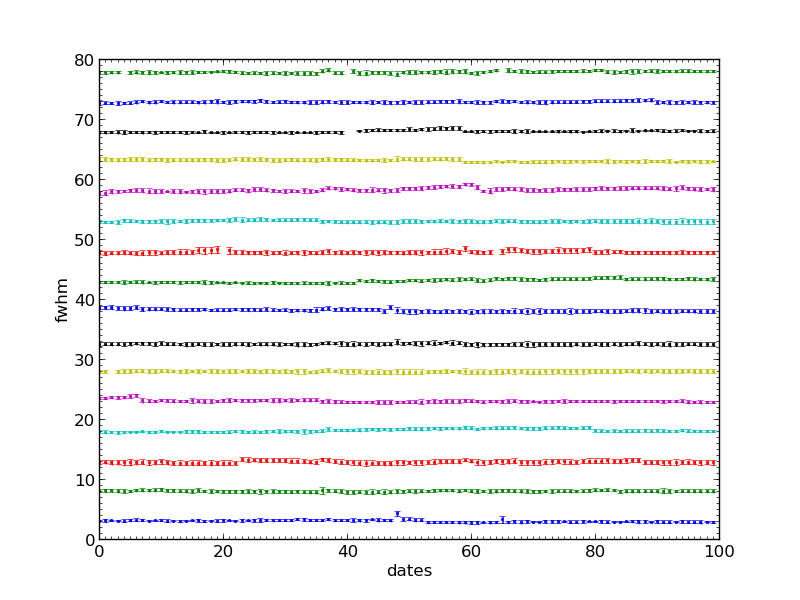}
\includegraphics[width=70mm]{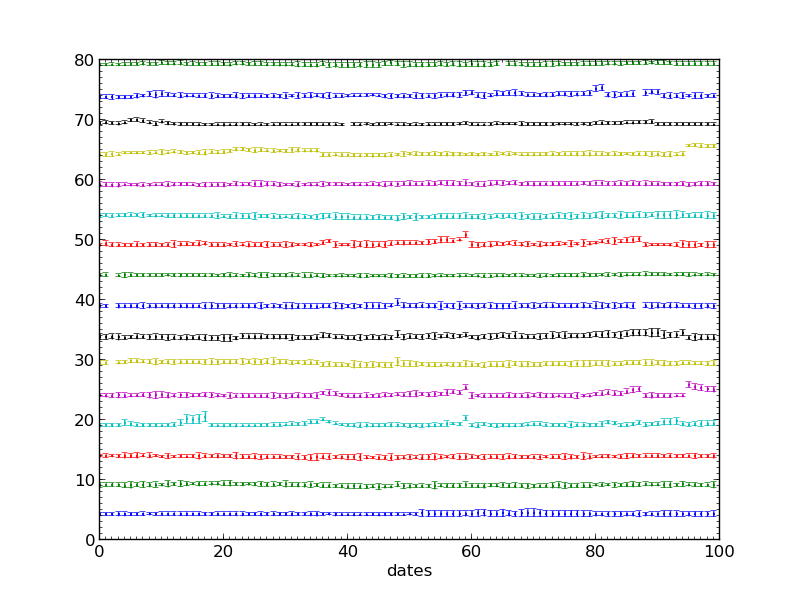}
\includegraphics[width=70mm]{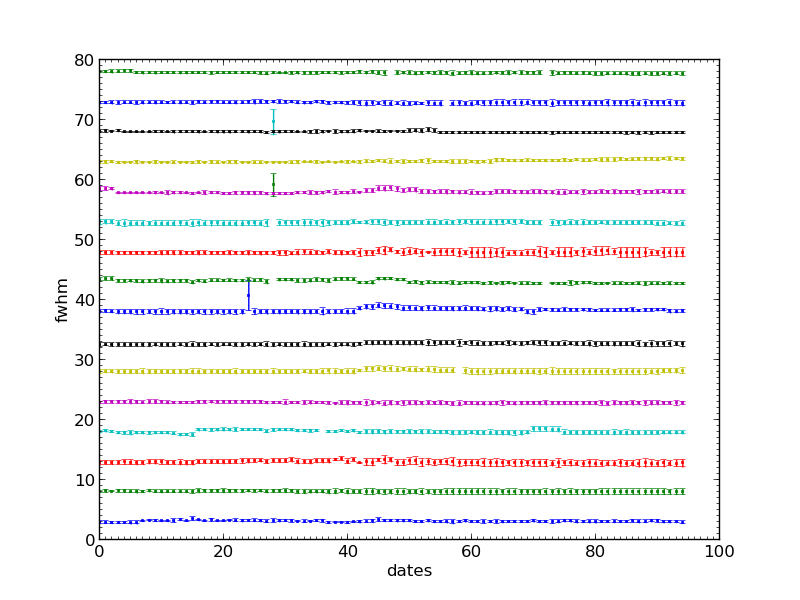}
\includegraphics[width=70mm]{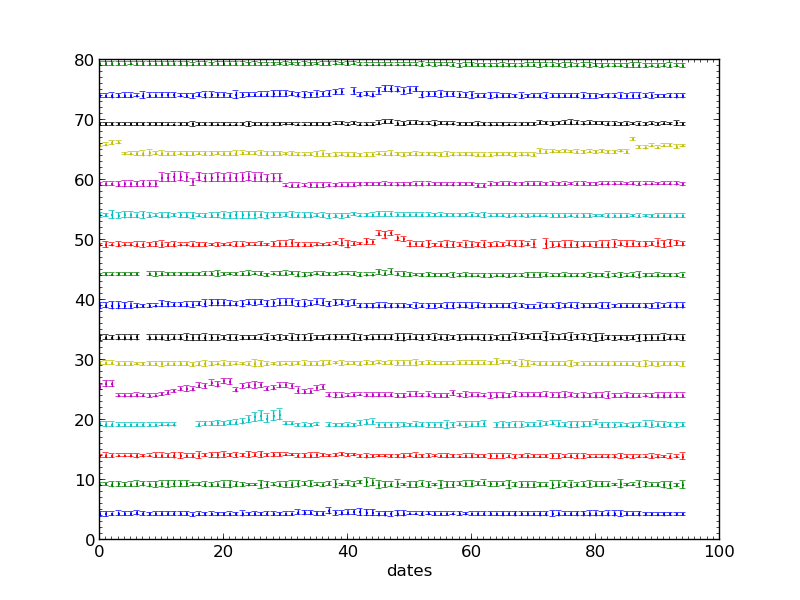}
\caption{{Variations of arc lines during continuous 300 days. The left panels show Full-Width-Half-Maximum (FWHM) of arc lines in the blue channel for the first 100, the second 100 and the last 100 days respectively, and the right panels are the same observations for the red channel.}\label{figure1}}
\end{figure}

\textit{Overhead} During the observations, the overhead time will reduce the observational efficiency of the survey to some extent. Specifically, the telescope needs to be confocal, guide stars need to be located, and the fibers need to be repositioned when a new plate is to be observed. Including the CCD readout time, the total overhead prior to observation of a field takes about 20 minutes. \\\\
\textit{Throughput} The throughput is a major indicator we use to measure the efficiency of the telescope. The whole throughput takes the contributions of the spectrograph, fibers, CCDs, optical system, etc., into account. In general, the throughput of the red and the blue channels are 4\% and 2\% respectively \citet{2012RAA....12.1197C}. During the pilot survey and the 1st year regular survey, the throughput of LAMOST was relatively low, and also showed large dispersion among the 4,000 fibers. The 4$^{th}$ spectrograph (out of a total of 16 spectrographs) has the highest throughput of those located near the center of focal plane; the median throughput of the valid fibers connected to the 4$^{th}$ spectrograph during the pilot survey was about 1.0\% and 2.0\% for the blue and red bands, and improved to 1.2\% and 3.0\% for the blue and red bands during the 1st year regular survey. The No. 3, 5, 8, 9, and 15 spectrographs are located at the middle-ring of the focal plane; their throughputs are about 80---90\% of that of the 4$^{th}$ spectrograph. The No. 1, 2, 6, 7, 10, 11, 12, 13, 14, and 16 spectrographs are located at the outer--ring of the focal plane, and have measured throughputs that are about 50---80\% of that of the 4$^{th}$ spectrograph. Among the 250 fibers accommodated by each spectrograph, the difference in the throughput can change by a factor of two.\\\\
\textit{Stability of instrument profile} Variations of emission lines from arc lamps have been monitored during a continuous 300 day period in order to measure the stability of the instrument profile. From Figure~\ref{figure1}, it can be seen that the profile of each fiber changes little with time, but that there differences among the profiles between the fibers. \\\\

\textit{Flux calibration.} The LAMOST project is designed as a spectroscopic survey without a photometry telescope equipped. Therefore, absolute flux calibration cannot be done without the photometric data \citep{2012RAA....12..453S}. Besides, not enough first--level standard stars are available over the large FoV observed by LAMOST that can be used for relative flux calibration. Thus we choose some stars with high quality spectra as standard stars. The uncertainty of the shape of the continuum in these standard stars will thus contribute uncertainty to the LAMOST flux calibration.

\subsection{Sky subtraction and flat fielding}
According to the requirements set for LAMOST observations, $\sim10-15\%$ of the fibers for each plate are assigned for modelling the sky. This fiber assignment strategy works well for fields at high Galactic latitude, providing sky-subtraction that is consistent with results from the SDSS survey. However, the proportion of the fibers that sample ``clean'' sky background is not adequate to build a supersky in observations near the Galactic disk, especially in bright nights. The emission lines of the night sky spectra are not subtracted satisfactorily for a number of spectra under these conditions. Meanwhile, the fiber-to-fiber profile differences makes the subtraction of sky emission lines more complicated. Moreover, the sky continua are more difficult to remove than narrow sky emission lines, such that there will always be at least some sky residual remaining in reduced spectra. This in turn presents a challenge we have to deal with when determining (stellar) parameters from spectra.

It is not easy to find a light source with a five-degree FoV to use as a flat field. We therefore separately correct the flat field for each spectrograph, since each spectrograph covers only about a 1$^\circ$ FoV. Twilight flats and screen flats are used. Additionally, the vignetting and the change of aperture caused by the telescope pointing also complicate the correction of the flat field. The movement of fibers may lead to focal ratio degradation caused by fiber tension changes, which will cause differences in the efficiency of fibers. 

\section{Data reduction}
\subsection{Statistics of data quality} 
\subsubsection{Resolution and re-sampling}
The theoretical resolution of the 16 LAMOST spectrographs is $R=1000$ (as determined by the gratings), and the designed wavelength coverage is 365--900~nm. In practice, the resolution reaches R$>$1500 by narrowing the slit at the output end of the fibers to 2/3 the fiber width. Since the spectral images are sampled on CCDs, to avoid confusion between resolution and sampling for the final re-binned spectrum, we explain the difference as follows. On the original CCD image, 4K pixels are used for recording blue or red wavelength regions ranging from 370~nm to 590~nm or from 570~nm to 900~nm respectively, which means each \AA~ has been sampled in 2 CCD pixels in the raw data. When the blue and red channels are combined together, each spectrum is re-binned for easier calculation of radial velocity. The combined spectra are re--sampled in constant-velocity pixels, with a pixel scale of 69~km~s$^{-1}$ , which means the wavelength difference between two adjacent points is $\Delta$ log ($\lambda$) = 0.0001. For convenience, the re-binned data interval is still referred to as 'pixel'.

\subsubsection{The definition of Signal to Noise Ratio}
\begin{figure}[!t]
\centering
\includegraphics[width=45mm]{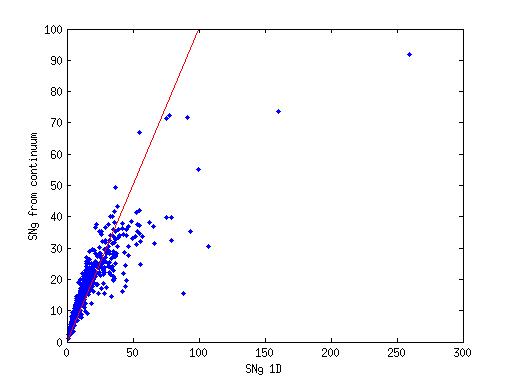}
\includegraphics[width=45mm]{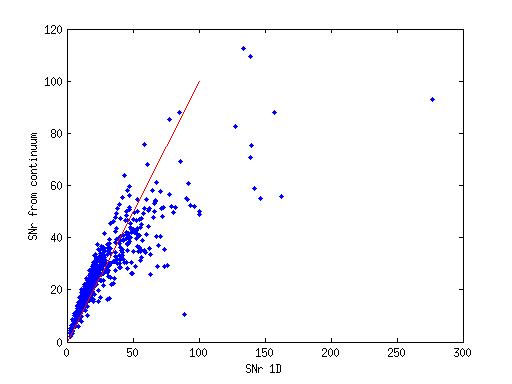}
\includegraphics[width=45mm]{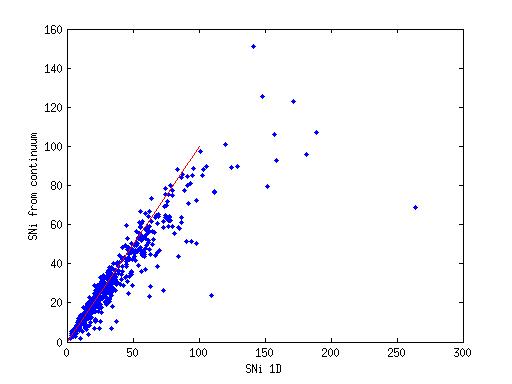}
\caption{{The two definitions of SNR are compared for a sample of LAMOST spectra. From left to right, the panels show SNR in the $g, r,$ and $i$ bands.}\label{figure2}}
\end{figure}

The signal to noise ratio (SNR) is always used as a quality indicator, and is the ratio of flux to noise in some bandpass. Generally, SNR is an average value in a wavelength band and represents predominantly the continuum flux level. SNR hereafter indicates the SNR {\it per 'pixel'}. In the LAMOST data products, SNR is defined by using the concept of inverse-variance, which is included in the spectral FITS files. It is easy to obtain the SNR for each pixel by calculating the product between the flux and its inverse-variance. The mean SNR in a wavelength band can be averaged from the SNR of each pixel in the band range.  Another way to obtain SNR is to calculate the ratio between the continuum and the 1${\sigma}$ standard variance.  These two definitions of SNR are shown to provide consistent results when applied to LAMOST data (Figure~\ref{figure2}).

\subsubsection{Data quality}
The data quality is determined by SNR, and a successful observational target should have SNR $>$10 in $g-$ or $i-$band. Table~\ref{tableX} lists the statistics of the spectrum success rate. The success rate improved from the pilot survey to the general survey, and the rate on VB plates reaches to 70\% even though these targets were observed in bright nights.  

\begin{table}
\begin{center}
\label{tableX}
\caption{The statistics of spectrum success rate}
\begin{tabular}[]{lrrr}
\hline
Plate type & Number    & Number of   	  & Rate of \\
	        & of plates & spectrographs   & successful spectra\\
\hline
GAC		&		190		&		2592		&		318854/531019 = 60\% \\
M31		&		41		&		511		&		54950/104770 = 52\% \\
VB			&			343		&		4405		&		552245/794100 = 70\%\\
HD		&		347		&		5124		&		479076/745791 = 64\% \\
EG		&		71		&		1057		&		102675/199091 = 52\% \\
F M B		&		202		&		2805		&		268154/570498 = 47\% \\
Kepler		&		8		&		104		&		15333/22899 = 67\% \\
Total		&		1202		&		16598		&		1791287/2968168 = 60\% \\
\hline
\end{tabular}
\begin{flushleft}
{\sc Notes:}\\
1. The last column is the spectrum success rate, which means the ratio of the number of spectra with SNR$_g$ or SNR$_r$ $\geq$ 10 to the number of fibers that were assigned to targets. \\
2. Plate types are named from the pilot survey and the first year of the general survey. The details are described as follows: \\
~~~~\textit{GAC}: ~~Galactic Anti--Center plates.\\
~~~~\textit{VB}: ~~~~Very bright stars with r$\leq$14 in all northern sky.\\
~~~~\textit{HD}: ~~~~Halo and disk plates (Sep. 2012---Jun. 2013).\\
~~~~\textit{EG}: ~~~~Extragalactic plates.\\
~~~~\textit{F M B}: Halo and disk plates(Oct. 2011---Jun. 2012). F, M, and B means faint, medium,  and bright, respectively.\\
\end{flushleft}
\end{center}
\end{table}

\subsection{2D pipeline}
\begin{figure}[h]
\centering
\includegraphics[width=140mm]{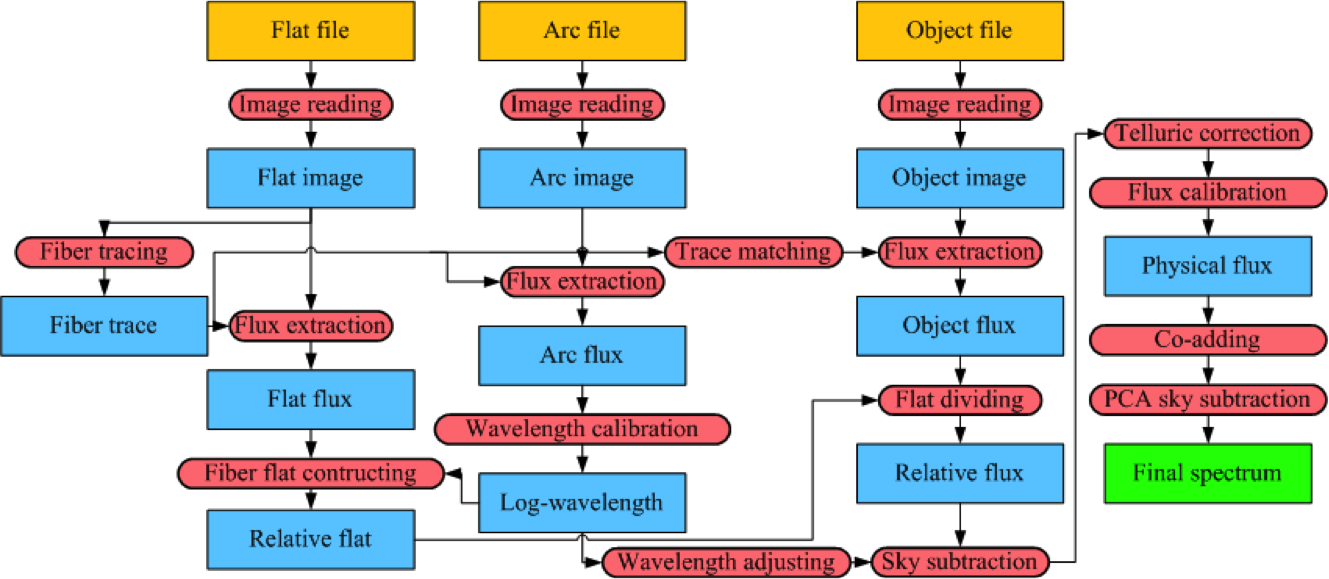}
\caption{{Flowchart illustrating the LAMOST 2D pipeline processing.}\label{figure3}}
\end{figure}

Raw CCD data were reduced by the LAMOST data reduction software known as the LAMOST 2D pipeline. The procedures of the 2D pipeline, similar to those of SDSS \citep{2002AJ....123..485S}, aim to extract spectra from CCD images and calibrate them. The main tasks of the 2D pipeline include dark and bias subtraction, flat field correction, spectra extraction, sky subtraction, wavelength calibration, sub-exposure merging and wavelength band combination. The processes of the 2D pipeline are listed as following (see the flowchart in Figure~\ref{figure3}).\\

1. Bias and dark counts are subtracted from each raw image. During the pilot survey, some inside CCD light source affected images, so that we needed the dark frames to eliminate this stray light. In the general survey, this light source was identified and removed, so that no dark subtraction is needed.\\

2. Flat field corrections are mainly achieved using twilight flat fields. However, white-screen flat fields were used on some days. LAMOST spectrographs are regarded as 16 different individual instrument systems for flat correction, so that the flat for each spectrograph is independent.\\ 

3. The flat-field spectra are traced for each fiber, and the centroid of the row position of each fiber is fitted by a polynomial function. A good fitting of each spectral track overcomes interference between fibers. However, the fitting will fail at the end of the wavelength coverage range when the SNR becomes low.\\

4. A spectrum is extracted along the track fitted by step 3 using a weighted aperture integration method; flux on 15 pixels is integrated to provide a single flux for 1D spectra.\\

5. The sky background is modeled and subtracted by a fixed proportion of the fibers on each plate that is reserved for pointing at blank sky. The method of B-spline fitting is applied to make a supersky for each sub-exposure and then Principal Component Analysis (PCA) is used to recorrect the supersky after combining in this procedure. Additionally, the relative strength of sky emission lines is used to correct efficiency differences between sky fibers and object fibers, since these lines exists both in sky spectra and object spectra.\\

6. Wavelength calibration is by use of the arc lines. The arc lamp spectra are extracted, and centroids of the lines are measured, to which we fit a Legendre polynomial as a function between wavelengths and pixels, with a fifth-order polynomial in the blue branch and sixth-order in the red. A vacuum wavelength scale is applied for wavelength calibration, is slightly adjusted to match the known positions of certain sky lines, and then corrected to the heliocentric frame. \\

7. The telluric absorption in four wavelength regions (6850---6960~\AA, 7150---7350~\AA, 7560---7720~\AA, 8100---8240~\AA) is removed.\\

8. The spectra are flux-calibrated by matching the selected flux standard stars with their matching templates of F-type stars. First, 250 spectra from one spectrograph are sorted in descending order of SNR. Stars with bad SNR or saturated fibers are abandoned. After normalizing the continuum of each spectrum, atmospheric parameters are obtained by matching a template using the Kurucz model. Then stars with 6000 K $\leq T_{\rm eff} \leq$ 7000 K are selected as standard stars, after visual inspection to confirm their correct identification. If the number of standard stars is less than 3, then the effective temperature will be extended from 5500--7500~K to obtain more standard stars.\\

9. For each object in all the exposures, the spectra from the red and blue spectrograph are combined by stacking the points with corresponding wavelengths using a B-spline function, with inverse-variance weighting. Outliers due to cosmic rays are removed and masked, then errors of the fluxes are estimated. \\

\subsubsection{Uncertainties in the wavelength calibration}
During the wavelength calibration, errors are most likely to be introduced when finding centroids of the emission lines of the arc lamp. In these emission lines, not all of the lines are single -- some observed lines are actually blends of two or three lines. Therefore, the fitting quality of a unimodal function for these mixtures may have large uncertainties. Statistically, the average error of wavelength calibration for the arc lamp is less than 0.02~\AA~, corresponding to about 1~km~s$^{-1}$. Generally, the calibration accuracy of blue spectra is better than that of the red one, but it degrades below 4000 \AA.

Another error of wavelength calibration is from heliocentric correction by wrongly using Beijing local MJM instead of International Standard Time. The difference between local MJM and International Standard Time is 480 minutes, which results in a random error of 0.36 km s$^{-1}$.

\subsubsection{Uncertainties in the relative flux calibration}
If there are no standard stars observed on one of the spectrographs (for a given plate), combined object spectra spanning the entire wavelength range cannot be produced. This is because the matching of the blue- and red-channel spectra relies on matching their flux in the wavelength region where they overlap.
%without flux calibration, because the blue and red spectra can not be combined for the total 250 spectra in one spectragraph.

The reddening calibration is not considered in standard stars selection. The extinction is negligible in high Galactic latitude areas. However, reddening will affect flux calibration severely in the low Galactic latitude areas, biasing the continuum shapes of resulting spectra in these high-extinction regions.
%a reddening difference lefts in the flux--calibrated spectra, owing to inconsistent reddening.

The flux of calibrated spectra from different exposures for each object may be not on the same scale. Hence before combining them, we adjust them onto a uniform scale, and then combine the spectra from different exposures by the method of B-spline fitting, which is similar to the SDSS pipeline.

\subsection{1D pipeline}
The LAMOST 1D pipeline works on spectral type classification and measurement of radial velocity (RV) for stars or redshift for galaxies and QSOs. By using a cross--correlation method, the pipeline recognizes the spectral classes and simultaneously determines the redshifts or radial velocities from the best fit correlation function. In this procedure, a low order polynomial is added while matching the observed spectra with templates, since the physical continuum can be difficult to detect because of the low resolution, reddening, and uncertainties in the flux calibration. Currently we employ the same templates as SDSS. However, the templates will be updated with LAMOST spectra for reductions after DR1. Finally, the pipeline produces four primary classifications, namely STAR, GALAXY, QSO, and UNKNOWN.

Measuring redshifts and recognizing galaxy and QSO spectra are difficult tasks since the pipeline does not work as well as for stellar classification. An additional independent pipeline was designed for galaxy recognition and redshift measurement after the initial 1D pipeline is run. It automatically identifies galaxies from the objects with confidence of classification below 80 or a target type as ``galaxy'' in the input catalogue. The redshifts of galaxies are measured through line centers; a Gaussian function with sigma of 1.5 times the wavelength step is applied to the spectra to eliminate noise before line centers are measured. The continua smoothed by a median filter are divided out to complete the normalization. Those data points that exceed 2${\sigma}$ of a normalized spectrum are selected as candidate emission lines, and a set of Gaussian functions are used to fit the lines. All the line centers are matched with line lists spaced by steps of 0.0005 in redshift ($z$). If most of the lines are matched successfully with heavily weighted lines such as $H_\alpha$, OII, $H_\beta$, OIII, or NII for emission line galaxies, or NaD, Mgb, CaII H, or CaII K for absorption galaxies, the spectrum is classified as galaxy, and the corresponding $z$ is the raw redshift of the spectrum. To detect mistakes in the redshift measurement, this redshift is compared with the results of the 1D pipeline, and the final redshifts are decided by experts.
For QSOs, the recognitions and measurements highly depend on visual inspection; a QSO catalogue of DR1 detections will be released by \citet{Ai..RAA..inpre}.

\subsection{LAMOST Stellar Parameter Pipeline - LASP}
LAMOST has carried out the pilot survey and the first year general survey, producing more than 2.2~million spectra, among which nearly 2~million were stellar spectra. Through statistical analysis of the stellar physical properties of this tremendous spectroscopic database spanning wide sky coverage, LAMOST will definitely encourage and actively promote our understanding on the structure, chemistry, and kinematics of the Galactic disk and the halo as well as for its origin and evolution history. Efficient parameterization software to accurately derive the fundamental stellar atmospheric parameters plays a key role for the scientific exploration. For this purpose, we developed and implemented the LAMOST stellar parameter pipeline (LASP, version 1.0). 

\subsubsection{Input Data and Strategy for LASP} 
LASP has been used to automatically determine the stellar parameters (effective temperature: $T_{\rm eff}$, surface gravity: $\log{g}$, metallicity: [Fe/H], and radial velocity: $V_r$) for the AFGK-type stellar observations from the LAMOST survey. The input data for LASP are the output FITS spectra produced by the 2D and 1D pipeline. For the DR1 data release, the fits file data version is v2.6.4 (version according to 2D and 1D pipeline updating status, for DR2 data release the version will be upgraded to v2.7.5 after some improvement), with a resolving power of $R\sim1800$ covering wavelengths $3800-9000$~\AA. The survey observing nights are generally divided into three categories: dark nights (8 days before and after the new moon), test nights (3 days around the full moon, used for instrumental testing), and bright nights (the remaining nights in a lunar cycle). The current criteria required for stars to be processed by the LASP are: $final \_ class$ is STAR, and $final\_subclass$ is spectral type A, F, G, or K, and $g$-band SNR of $SNR_g \geq 15$ and $SNR_g \geq 6$ for the bright and dark nights, respectively. The $final\_class$ and $final\_subclass$ information is retrieved from the 1D pipeline results. 

\subsubsection{ Methods of the LASP} 
The LASP consecutively adopts two methods -- CFI and UlySS \citealt{2011A&A...525A..71W} -- to determine the stellar parameters. The CFI method is used to give a set of coarse measurements that serve as initial guesses for ULySS, which then gives the final measurement that is issued by LASP.

\textit{1. Correlation Function Interpolation (CFI) method:}
Initial stellar parameter estimates are provided by a method called the CFI, short for Correlation Function Interpolation \citep{2012SPIE.8451E..37D}. The core algorithm of the CFI method is as follows: provided the observed flux vector is $O$, and the synthetic flux vector is $S$, the best-fit case is $\cos {< O, S >} = 1$ theoretically. The algorithm searches for the best-fit by maximizing $\cos {< O, S >}$ as functions of $T_{\rm eff}, \log{g}$, and [Fe/H], where $\cos{<O,S>} =\frac{O\cdot S}{\left | O \right |\times \left | S \right |}$, which is also referred to as correlation coefficient. Practically, CFI first explores in temperature among the grid of synthetic spectra to determine the $T_{\rm eff}$ value of the input spectrum. After $T_{\rm eff}$ is fixed, the code searches for the [Fe/H] in metallicity space among the grid, and then with fixed $T_{\rm eff}$ and [Fe/H], it searches for the $\log{g}$. According to maps of the correlation coefficient distribution for the 3 parameters, the reliability for $T_{\rm eff}$ estimation is relatively more credible, the [Fe/H] is less credible, and it is hard for the algorithm to determine a $\log{g}$ solution. Thus the biweight mean of several $\log{g}$ estimations with top correlation coefficient values is given out as the output.       

For the production of the synthetic spectra grid used by our CFI algorithm, we employed Kurucz spectrum synthesis code based on the ATLAS9 stellar atmosphere models provided by Castelli \& Kurucz (2003). This grid is smoothed to the resolution of the LAMOST spectrograph ($R\sim1800$) and the pseudo-continuum of the grid is manually obtained by an experienced spectral analyser. The analyser manually marks the points which may be on the pseudo-continuum based on spectral features. Then spline curves are employed to fit the spectral pseudo-continuum. Finally, the points can be adjusted until the pseudo-continuum meet the requirements. The CFI's adopted synthetic library contains 8903 spectra, in a grid with stellar parameter coverage of: $3600 K \leq T_{\rm eff} \leq 7500 K$ in steps of 200~K, $0.0 {\rm dex} \leq \log{g} \leq 5.0$~dex in steps of 0.25 dex, and  $-3.0 \leq$~[Fe/H]~$\leq 0.4$ in steps of 0.2~dex.\\

\textit{2. ULySS method:}
The ULySS (Universite de Lyon Spectroscopic analysis Software, \citep{2009A&A...501.1269K,2011A&A...525A..71W}, available at \url{http://ulyss.univ-lyon1.fr/} package is employed by LASP to derive the stellar parameters via minimizing the squared difference between the observations and the model. The model is adjusted at the same resolution and sampling as the observation, and the fit is performed in the pixel space. This method determines all free parameters within a single fit in order to properly handle the degeneracy between some parameters, e.g., temperature and metallicity. \citet{2011A&A...525A..71W} have already described the ${\chi}^{2}$ minimization of the ULySS. The parametric model is: 

\begin{eqnarray*}
Obs(\lambda )=P_{n}(\lambda )\times[TGM(T_{\rm eff}, \log{g}, {\rm [Fe/H]}, \lambda )\otimes G(\upsilon_{sys}, \sigma)]
\end{eqnarray*}

where $\lambda$ is the logarithm of the wavelength, $Obs(\lambda)$ is the observed stellar spectrum sampled in $\log{\lambda}$, $P_n(\lambda)$ a series of Legendre polynomials of degree $n$, and $G(\upsilon _{sys}, \sigma)$ is a Gaussian broadening function characterized by residual velocity $\upsilon_{sys}$, and the dispersion $\sigma$. The $\sigma$ encompasses both the instrumental broadening and the effects of rotation.The multiplicative polynomial is used to absorb errors in the flux calibration, Galactic extinction or any other source affecting the shape of the spectrum. It replaces the prior rectification or normalization to the pseudo-continuum that other methods require. The parameters of the TGM are evaluated together with those of the LOSVD with a Levenberg-Marquardt \citep{marquardt1963algorithm,more1980user} routine (hereafter LM). The (linear) coefficients of the $P_n(\lambda)$ polynomials are determined by ordinary least-squares at each evaluation of the function minimized by the LM routine. The gaps and bad pixels (e.g. emission lines, cosmic ray, bad sky subtraction, telluric lines or bad ones due to the instrument errors etc.) are automatically rejected during the fitting process by ULySS, using the clipping algorithm applied iteratively on the residuals to the fit.

The non-linear TGM function is an interpolator that can return a spectrum for a given set of atmospheric parameters [$T_{\rm eff}, \log{g}$, [Fe/H]] by making an interpolation over the whole ELODIE library \citep{2001A&A...369.1048P, 2004astro.ph..9214P,2007astro.ph..3658P}. The interpolator (available at \url{http://
ulyss.univ-lyon1.fr/models.html}) consists of polynomial expansions of each wavelength element in powers of $\log{T_{\rm eff}}, \log{g}$, [Fe/H], and $f(\sigma$) (a function of the rotational broadening parameterized by $\sigma$, the standard deviation of a Gaussian). Three sets of polynomials are defined for three temperature ranges (roughly matching OBA, FGK, and M types) with important overlap between each other where they are linearly interpolated. For the FGK and M polynomials, 26 terms are adopted; for OBA, 19 terms are used. The coefficients of these polynomials were fitted over the $\sim$ 2000 spectra of the library, and the choice of the terms of the $T_{\rm eff}$ limits and of weights were fine tuned to minimize the residuals between the observations and the interpolated spectra. This interpolator built on the ELODIE library (version 3.2) provides valid inverted atmospheric parameters covering $3100 \sim 59000$~K in $T_{\rm eff}$, $0.00 \sim 5.00$~dex in $\log{g}$, and $-2.80 \sim 1.00$~dex in [Fe/H], from O to M type stars. The derived intrinsic external accuracies of ULySS for high quality AFGK stellar spectra are 43~K, 0.13~dex, and 0.05~dex for $T_{\rm eff}$, $\log{g}$, and [Fe/H] respectively \citep{2011A&A...525A..71W}. The performances of ULySS applied to LAMOST commissioning period observations were already illustrated in \citet{2011RAA....11..924W}.

\subsubsection{Processing Procedure of the LASP}

LASP operates in two stages to measure the stellar parameters. Since the LAMOST spectral flux calibration is relative (because there is no valid complete system for photometric standard stars), in the first stage we measure the original spectrum, and in the second stage we use the normalized spectrum. For the medium resolution of the LAMOST survey spectra, the blue band spectra contain most of the physical features necessary to constrain the stellar parameters. To avoid the effects of low instrumental response near both wavelength edges, as well as for optimizing the computing time and using minimal storage space, CFI selects wavelengths of $3850-5500$~\AA, and ULySS selects $4400-5700$~\AA (in the 1st stage) and $4400-6800$~\AA (2nd stage) for the fitting range. 

\begin{figure}[h]
\centering
\includegraphics[width=60mm]{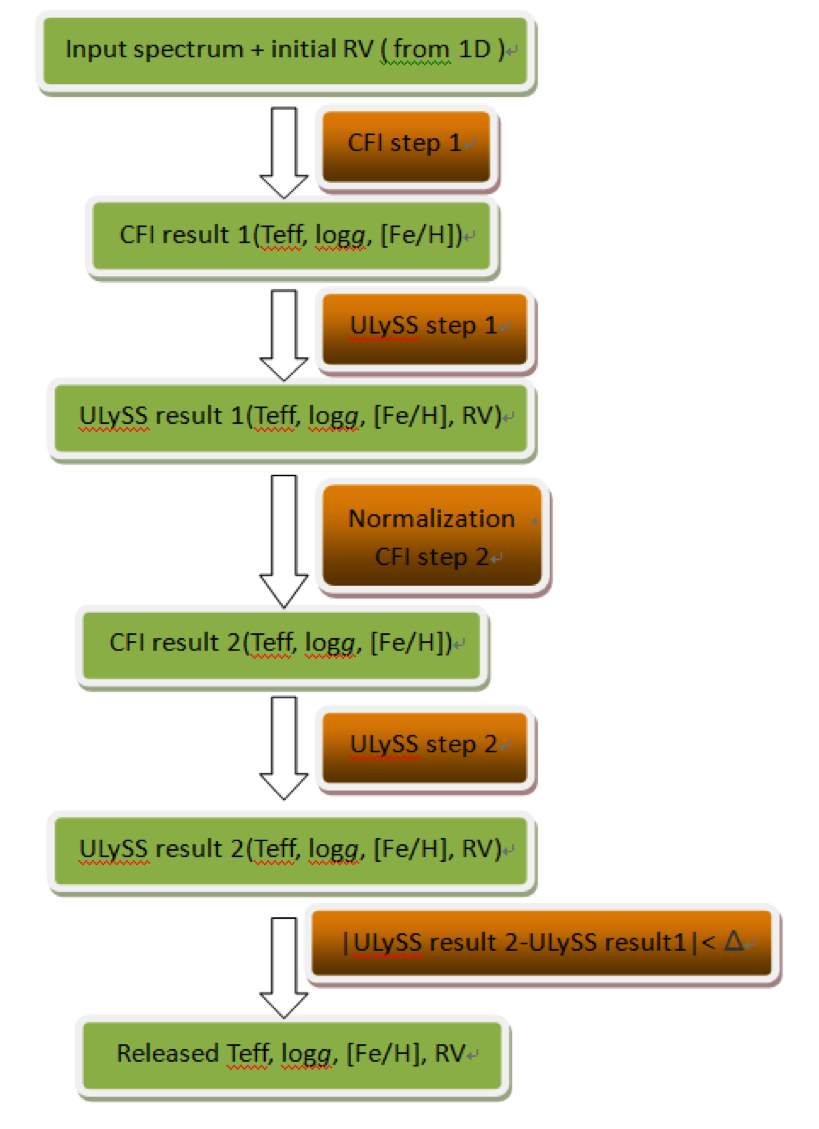}\\ 
\caption{{Flow chart illustrating the parameter estimation process of LASP.}\label{figure4}} 
\end{figure}

In each stage, we first utilize CFI to get a set of initial coarse estimates. After this, we use CFI results as starting guesses for ULySS to obtain more accurate and credible measurements. Figure~\ref{figure4} displays a general flowchart of the LASP. 

\begin{figure}[h]
\centering
\includegraphics[width=70mm]{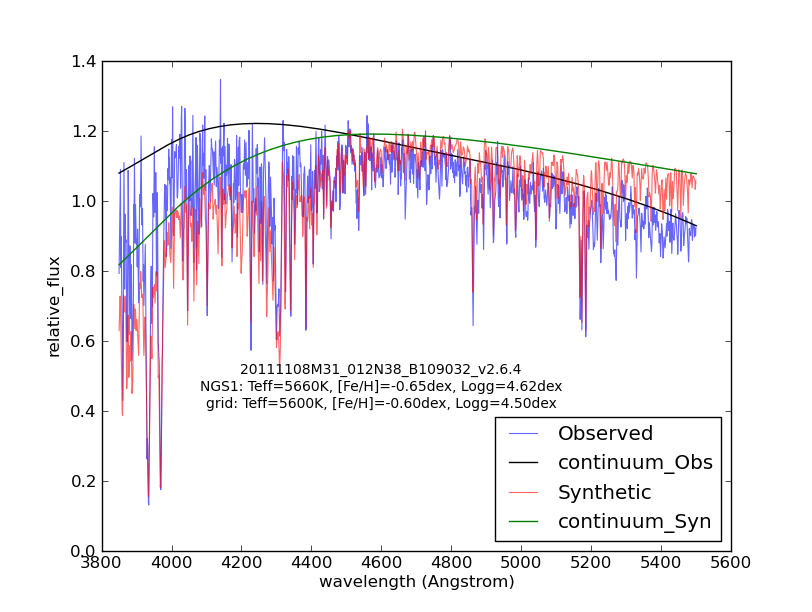} 
\includegraphics[width=70mm]{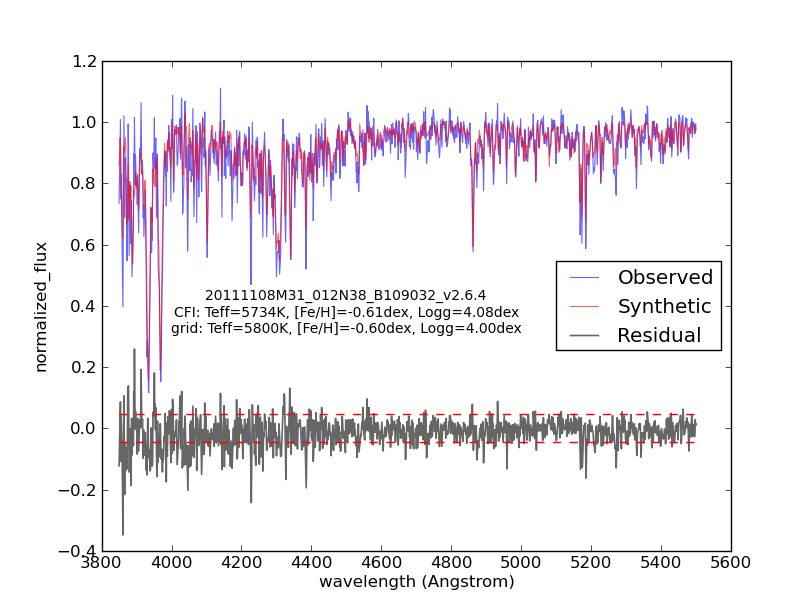}
\caption{{An example of normalization through the CFI grid. Left: The pseudo-continuum derived from the corresponding parameter grid. Right: The normalized flux of the blue wavelength range (in blue) and the synthetic grid spectrum (in red), showing that the spectrum has been well fit by the CFI method.}\label{figure5}}
\end{figure}

As CFI runs its core algorithm to determine the parameters individually one after another, it determines the best match between the observation and the synthetic grid spectra, both of which have been normalized. In the first stage of CFI, it uses the radial velocity (RV) estimated by the 1D pipeline to shift the observed spectrum to the rest frame. Theoretically, observed spectra could be approximated by synthetic spectra multiplied by low order polynomials. To quickly get an initial guess of the effective temperature for CFI as a prior, the code uses the NGS1 \citep{2008AJ....136.2022L} grid. The NGS1 grid of synthetic spectra was generated from the Kurucz NEWODF models, which employ solar relative abundances from Grevesse \& Sauval (1998). More details about generation of NGS1 synthetic spectra were described in \citet{2008AJ....136.2022L}. These synthetic spectra were convolved to $R = 1000$, then sampled into 1.67~\AA~per pixel to be the NGS1 grid spectra. In order to speed up the analysis, only the spectral range of 4500~\AA~to 5500~\AA~was used. The synthetic spectra have been normalized by using a 9th order polynomial, iteratively rejecting points that are more than 1$\sigma$ below and 4$\sigma$ above the fitted function. The NGS1 grid covers 4000~K$\leq$T$_{\rm eff}\leq$9750~K in steps of 250~K, 0.0~dex$\leq \log{g} \leq$5.0~dex in steps of 0.25~dex, and $-4.0 \leq$~[Fe/H]~$\leq +0.5$~dex in steps of 0.25 dex for the three parameters respectively. The observed LAMOST spectra are also processed in the same fashion over the same wavelength range, then we search for the best-fit using the CFI core algorithm to get an initial guess of $T_{\rm eff}$.  Then with this $T_{\rm eff}$, CFI uses its core algorithm to search for the most similar synthetic spectra (with already known pseudo-continuum, which could be subtracted before fitting) in the grid described before in Sec.4.4.2.  We use a 5th order polynomial to absorb the difference between the normalized observation and the normalized synthetic spectra, and finally get an estimation of the three parameters one after another. The grid described in Sec. 4.4.2 is more intensive than the one of NGS1. In the second stage of CFI, it takes the ULySS first stage results [$T_{\rm eff}$, $\log{g}$, [Fe/H], RV], which provide a more accurate RV to correct to the rest frame, and removes the need to use the NGS1 grid, since we can use the $T_{\rm eff}$ given by the ULySS 1st result. Since the spectra dealt with in the second stage are normalized, CFI directly matches the spectra without a prior step removing the pseudo-continuum. Figure~\ref{figure5} shows an example of the CFI spectral fitting both in the blue range and whole range. 

\begin{figure}[h]
\centering
\includegraphics[width=100mm]{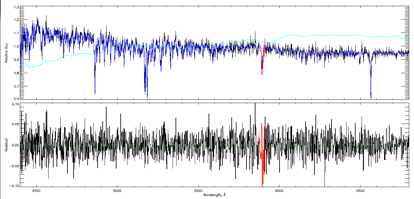}
\caption{{Fits of a LAMOST stellar spectrum with the ULySS ELODIE interpolator. The top panel represents the flux distribution -- the observation is in black, and the model is in blue. The light blue is the multiplicative polynomial. Red lines represent the masked NaD telluric lines which were not well calibrated in ELODIE. The bottom panel shows the residuals between the observation and the best-fitted interpolated spectrum (O-M). This spectrum shows star specid=20111108M31\_012N38\_B109032\_v2.6.4 with $g$-band SNR=17.51. The CFI\_stage1 parameters are: [$T_{\rm eff}, \log{g},$ [Fe/H]] = [5734.07 K, 4.084 dex, -0.61 dex]; CFI\_stage2 results: [5690.11 K, 4.01 dex, -0.74 dex], ULySS\_stage1 results: [5546.84 K, 4.31 dex, -0.40dex, -45.0 km/s], and ULySS\_stage2 results: [5605.01 K, 4.37 dex, -0.34 dex, -42.5 km/s].}\label{figure6}}
 \end{figure}

Before stellar spectra are parameterized, the pseudo-continuum must be divided out (or normalized).  For the input observation, its analogous synthetic spectrum can be found in the synthetic grid (Sec 4.4.2) according to the parameters of ULySS 1st stage results. Since the pseudo-continuum of the grid spectra have been derived previously, the pseudo-continuum of the observed spectrum can be calculated through the analogous synthetic spectrum multiplied by a polynomial. In order to obtain a continuum fit over the whole range, the wavelength range is divided into four pieces ($\leq$ 5600 \AA, 5400 - 6400 \AA,  6200 - 8000 \AA, $\geq$ 7800 \AA) and a 5th order polynomial is applied to each of these ranges. After this piece-wise fitting, the pseudo-continuum of the entire wavelength range is fitted by a 11th order polynomial, then the original 1D spectrum is normalized and delivered for the second stage. We made experiments of this normalization technique on the well calibrated SDSS/SEGUE \citep{2000AJ....120.1579Y,2009AJ....137.4377Y} survey spectra, and found that the difference between the determined parameters given by ULySS before and after normalization is negligible. However, when we applied this method to the LAMOST observations the differences are noticeable. This difference is mainly due to the unstable relative flux calibration of the spectra mentioned in Sec. 4.2.2; sometimes the shape of the flux-calibrated spectrum is quite strange, for example showing very steep regions or with several big fake humps or other spurious features, which obviously cannot be fitted by a fixed order polynomial. On the other hand, this simple intervening normalization procedure can potentially alter of shape of the original physical lines. So besides the $g$-band SNR, the difference in the ULySS parameters between first and second stage results, to some extent, helps reflect the spectral data quality.
% The reasonable degree of approximation can be measured through statistics. 
If the two stages of ULySS measurements are similar to each other, ULySS determinations given by the normalized spectra are provided in the final publicly released AFGK stellar parameter catalog. Figure~\ref{figure6} illustrates one fitting example of ULySS for LAMOST observations.

For observations with no prior knowledge, in order to prevent the ULySS algorithm from becoming trapped in a local minimum, we use a grid of initial guesses sampling all of the parameter space. For example, take the nodes of the guessing grid as $T_{\rm eff}$ =[4000, 7000, 15000] K, $\log{g}$=[1.8, 3.8] dex, and [Fe/H] = [-1, 0] dex as adopted in the work of \citet{2011RAA....11..924W}. The final solution (absolute minimum) is the best from those obtained with different guesses. For high quality AFGK type spectra, statistically by using a wide guessing grid, the global solution could be almost always be successfully found, with only several cases where the solution was a local minimum occurring out of ten thousand fitted stars. Also, for some spectra, the solution simply expressed by three physical parameters may not be unique. ${\chi}^{2}$ maps together with the convergence maps tool of the ULySS package (see \citep{2011A&A...525A..71W}) were used to visualize the parameter space and identify the degeneracies and local minima. We can see that if taking the above guessing grid, each observation has to be fitted 12 times. For a survey project with massive data flow, it is quite time consuming, which is why we use CFI in front of ULySS to get a single initial guess. This effectively helps reduce the computational needs required, saving time and storage space for LASP by using fewer starting guesses in ULySS.

Though CFI is just used to give a coarse estimation, for most cases its results used for the initial guesses in ULySS are acceptable ($T_{\rm eff}$ is most reliable). In some cases, due to its low accuracy related either to the algorithm (those parameters are not physical isolated, or the spectral SNR if poor, for example) or from poor parameter coverage of the synthetic spectral grid, some of the determinations given by CFI are not credible. These biases are obvious for very hot and cold stars, and also for wide regions in the $\log{g}$ and [Fe/H] spaces. In practice, if any of the CFI parameters reside around the edges of its grid parameter coverage, or in an unphysical region, or out of the grid, ULySS does not take its results, but instead either uses a moderate guess like [7000 K, 3.0 dex, 0.0 dex] or roughly estimates the starting value for $T_{\rm eff}$ from the subclass estimated by the 1D pipeline.
%classification (currently is not very precise). 

Note that for the LAMOST DR1 data release, if the star is processed by LASP (i.e., it has $T_{\rm eff}, \log{g}$, and [Fe\/H] values), the final released RV is determined by ULySS. If the spectrum was not processed by LASP, the RV is that measured by the 1D pipeline, and for late M-type stars, the RV is from a separate module (inside the 1D pipeline) which is specially tailored for cooler stars \citep{Guo..RAA..inpre}.

\subsubsection{Why iterate once}
As known from the normalization described in section 4.4.3, the quality of the calculated pseudo-continuum depends on the stellar atmospheric parameters estimated by ULySS. There are large discrepancies between the synthetic and observed spectrum if the stellar atmospheric parameters from ULySS are unreliable. Therefore, the pseudo-continuum of observed spectrum calculated based on the synthetic spectrum is unacceptable, and the stellar atmospheric parameters obtained by ULySS using normalized spectra and parameters using un--normalized spectra will not agree. Likewise, if the two stellar atmospheric parameter estimations agree, it can be concluded that the stellar atmospheric parameters are reliable and the pseudo-continuum of the observed spectrum calculated based on the synthetic spectrum is acceptable. The reasonable degree of approximation can be measured statistically. In brief, if the two results approximately agree with each other, the stellar atmospheric parameters estimated using normalized spectra are adopted.

\subsubsection{Errors on the Stellar Atmospheric Parameters of LASP}

Since reliable random errors for each wavelength element are unavailable for LAMOST spectra, like in the work of \citet{2011A&A...525A..71W}, when using ULySS, we assumed a constant noise at all wavelength points. We estimated an upper limit to the internal errors on the derived parameters by assuming  ${\chi}^2 =1$. We performed the fit with an arbitrary value of SNR and re-scaled the internal computed errors returned by ULySS by multiplying them by ${\chi}^2$. The computed internal errors are small, likely for reasons related to the internal degeneracies between the parameters. We used statistics from comparison of the common targets observed by both LAMOST and SDSS/SEGUE \citep{2014IAUS..298..445W,2014arXiv1407.1980W} to estimate the external error. We then rescaled the errors using the ratio of the differences between LAMOST and SDSS parameters to the formal errors, assuming that the uncertainties of each series are equivalent. Currently the external errors determined in this way released in the parameters catalog are larger than the real measurement errors. 

\subsubsection{Internal Comparison}

\begin{figure}[h]
\centering
\includegraphics[width=55mm]{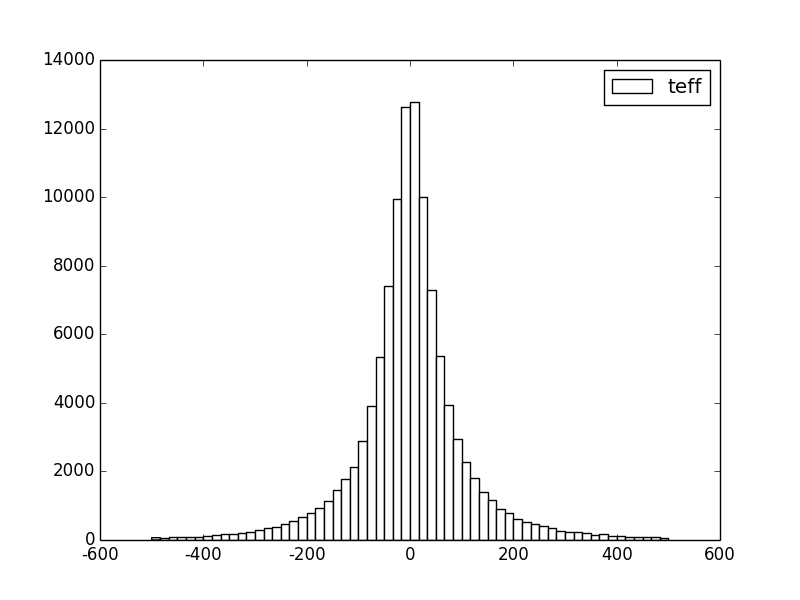}
\includegraphics[width=55mm]{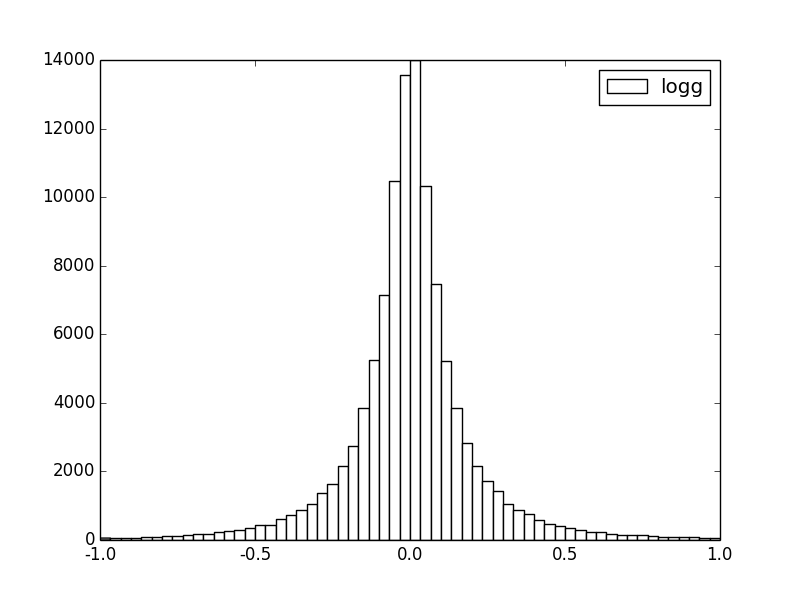}
\includegraphics[width=55mm]{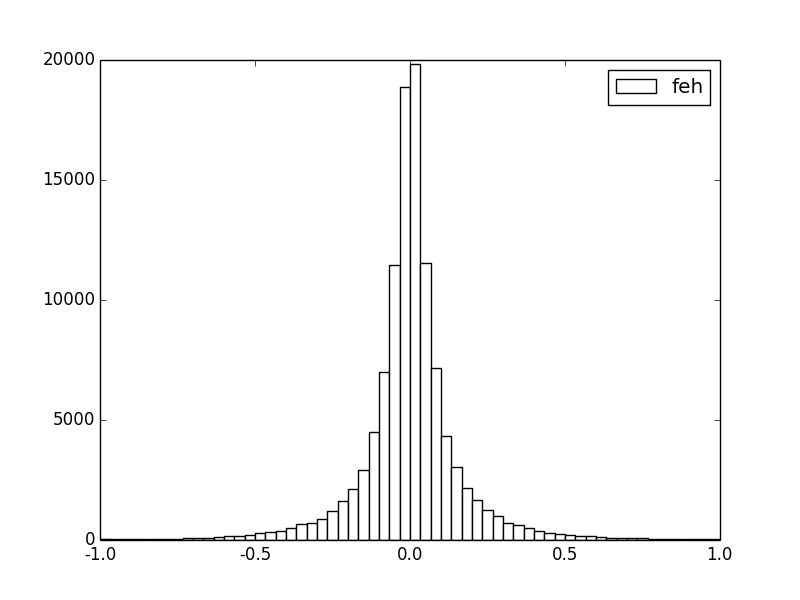}
\includegraphics[width=55mm]{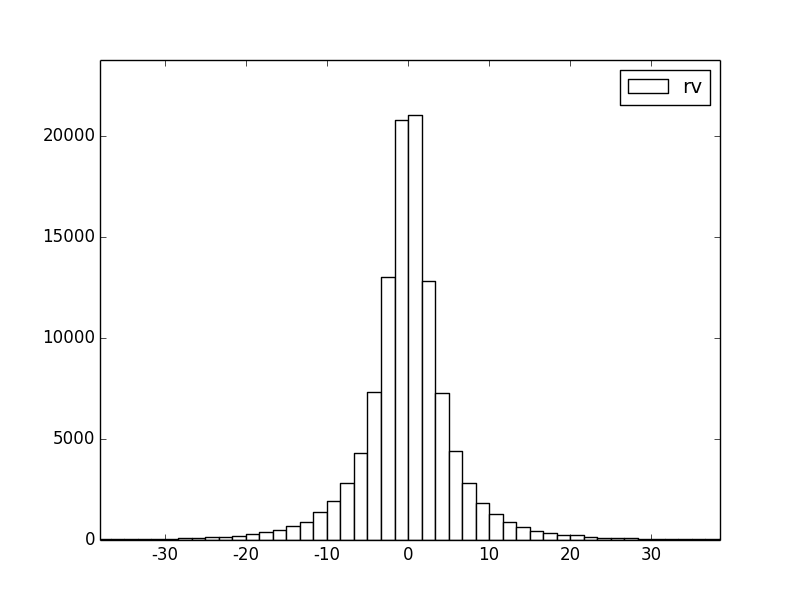}
\caption{{Histograms comparing differences in measured stellar parameters for pairs of observations of the same star. The panels show the internal consistency of $T_{\rm eff}$, $\log{g}$, [Fe/H], and RV for 111,000 targets.}\label{figure7}}
\end{figure}

In total, about 18\% of the targets in LAMOST have been observed more than one time. These multiple exposures can be used to estimate the internal errors on the parameters. We selected 111,000 targets with two observations to estimate the internal errors. Figure~\ref{figure7} demonstrates the internal error distribution (i.e., the difference between parameters measured for the same star from different spectra) including a comparison of $T_{\rm eff}, \log{g}$, [Fe/H], and RV.

\begin{figure}[h]
\centering
\includegraphics[width=55mm]{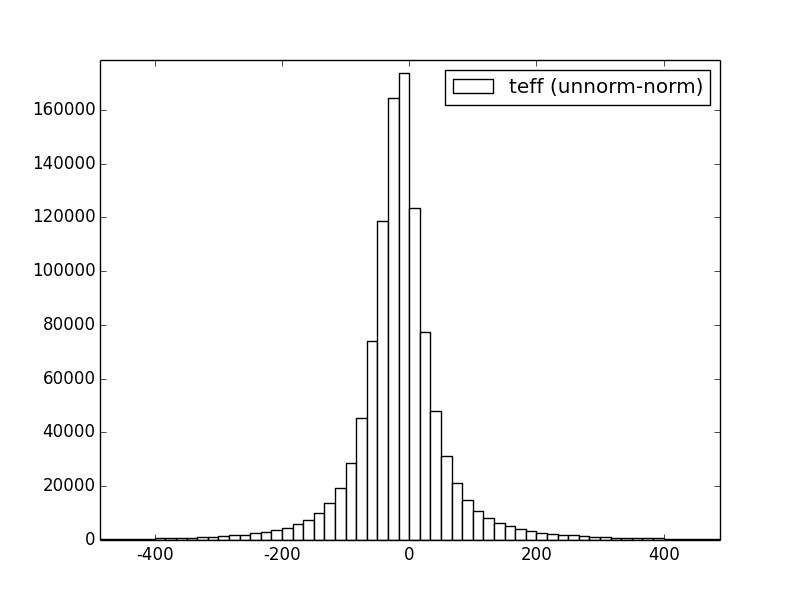}
\includegraphics[width=55mm]{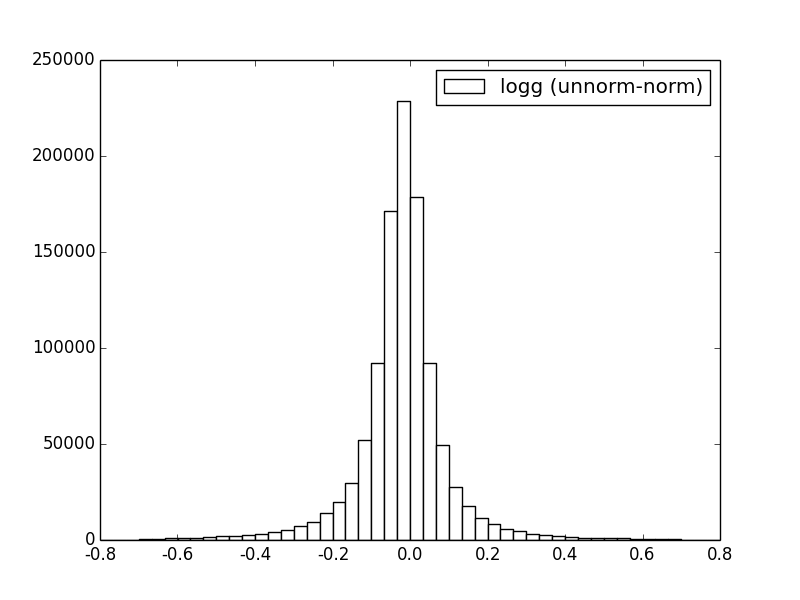}
\includegraphics[width=55mm]{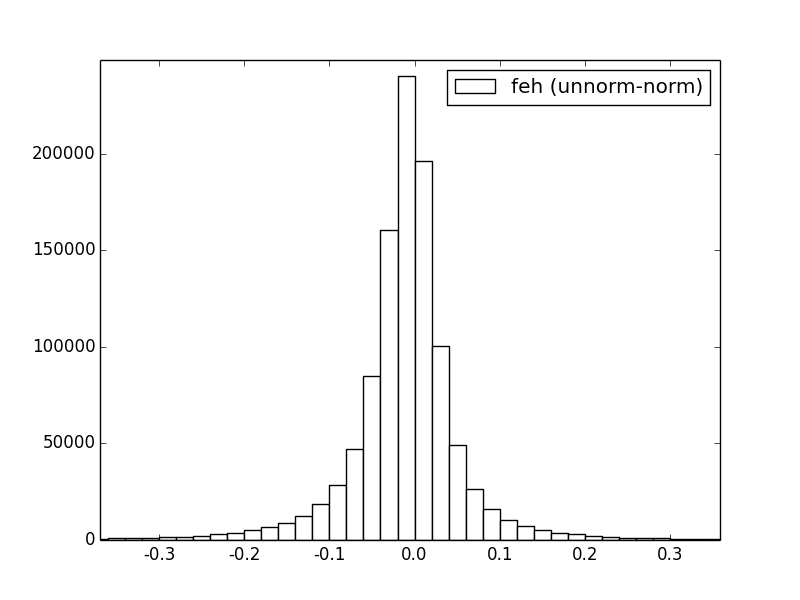}
\includegraphics[width=55mm]{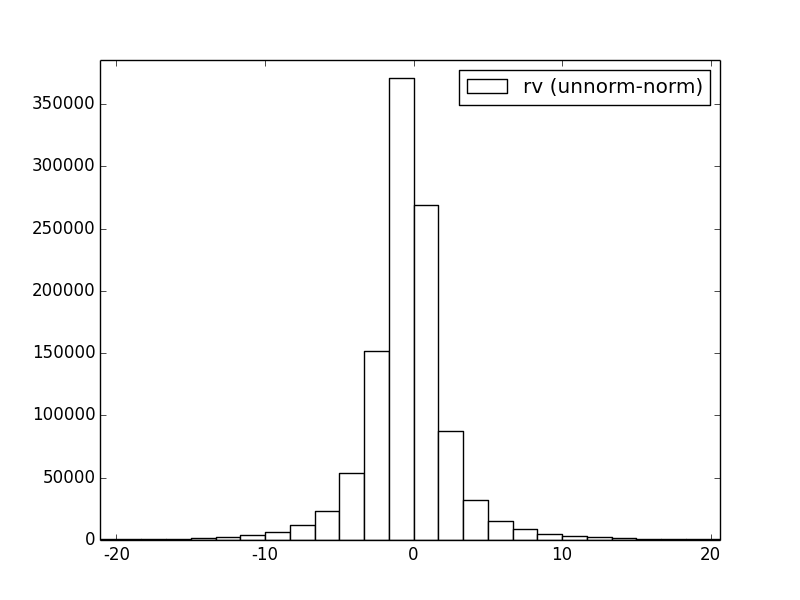}
\caption{{As in Figure~\ref{figure7}, but now showing the difference between parameters measured using normalized and un-normalized spectra.}\label{figure8}}
\end{figure}

Another way to estimate the internal error is comparing the parameters calculated through normalized spectra and un-normalized spectra. Figure~\ref{figure8} shows the distribution of difference between the parameters measured using normalized and un-normalized spectra.

\subsubsection{External errors}
\begin{figure}[h]
\centering
\includegraphics[width=105mm]{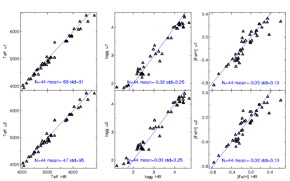}
\caption{{Comparison of the three parameters measured by LASP on LAMOST DR1 spectra and results from high resolution (HR) spectra (or Kepler asteroseismology, in the case of $\log{g}$) of the same stars. $u_1$ means the ULySS determination from the original spectra, and $u_2$ means the ULySS determination from the normalized spectra.}\label{figure9}}
\end{figure}

\begin{figure}[h]
\centering
\includegraphics[width=120mm]{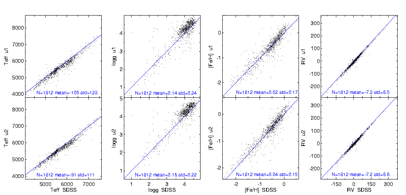}\\ 
\caption{{Comparison of the three parameters between LASP and SDSS DR9 . The convention is same as described in Figure 9.}\label{figure10}}
\end{figure}

We compared the common targets that have both LAMOST DR1 LASP measurements and high resolution spectral results \citep{2012MNRAS.423..122B,2012A&A...543A.160T,2013MNRAS.434.1422M, 2014ApJ...789L...3D}. For this comparison, $T_{\rm eff}$ and [Fe/H] come from the high resolution spectra, while $\log{g}$ comes from Kepler \citep{2010Sci...327..977B} asteroseismic estimation. The comparison between LAMOST DR1 and high-resolution data is shown in Figure~\ref{figure9}, along with statistical fits characterizing the quality of agreement. 
Figure~\ref{figure10} shows a similar comparison for the common targets between LAMOST (LASP results) and SDSS DR9.

There are several independent works comparing parameters of DR1 with other reliable external databases. For example, \citeauthor{Gao..RAA..inpre} compare DR1 parameters with the PASTEL catalog. Results from LSP3 \citep{2015MNRAS.448..822X}, an independent parameter pipeline developed at PKU, shows very good consistency with LAMOST/LASP measurements. All of these comparisons are summarized in Table~\ref{table2}. 

\begin{table}
\begin{center}
\caption{{External Comparison with DR1}\label{table2}}
\begin{tabular}[]{rlllr}
\hline
~ &~$T_{\rm eff}$ error & ~~$\log{g}$ error & ~~~[Fe/H] error & rv error~~~~~~~~~~\\
\hline
High resolution &-47$\pm$95 K & 0.03$\pm$0.25 dex & -0.02$\pm$0.1 dex &--~~~~~~~~~~~\\
SDSS DR9       &-91$\pm$111 K &0.16$\pm$0.22 dex &0.14$\pm$0.05 dex &-7.2$\pm$6.6 km/s\\
Gao et al.    & ~11$\pm$110 K & 0.03$\pm$0.19 dex & ~0.01$\pm$0.11 dex & -3.78$\pm$4.91 km/s \\
LSP3    &~~7$\pm$133 K & 0.02$\pm$0.22 dex & ~0.06$\pm$ 0.14 dex  & 1.2$\pm$ 3.1 km/s\\
\hline
\end{tabular}

\end{center}
\end{table}

\begin{figure}[h]
\centering
\includegraphics[width=120mm]{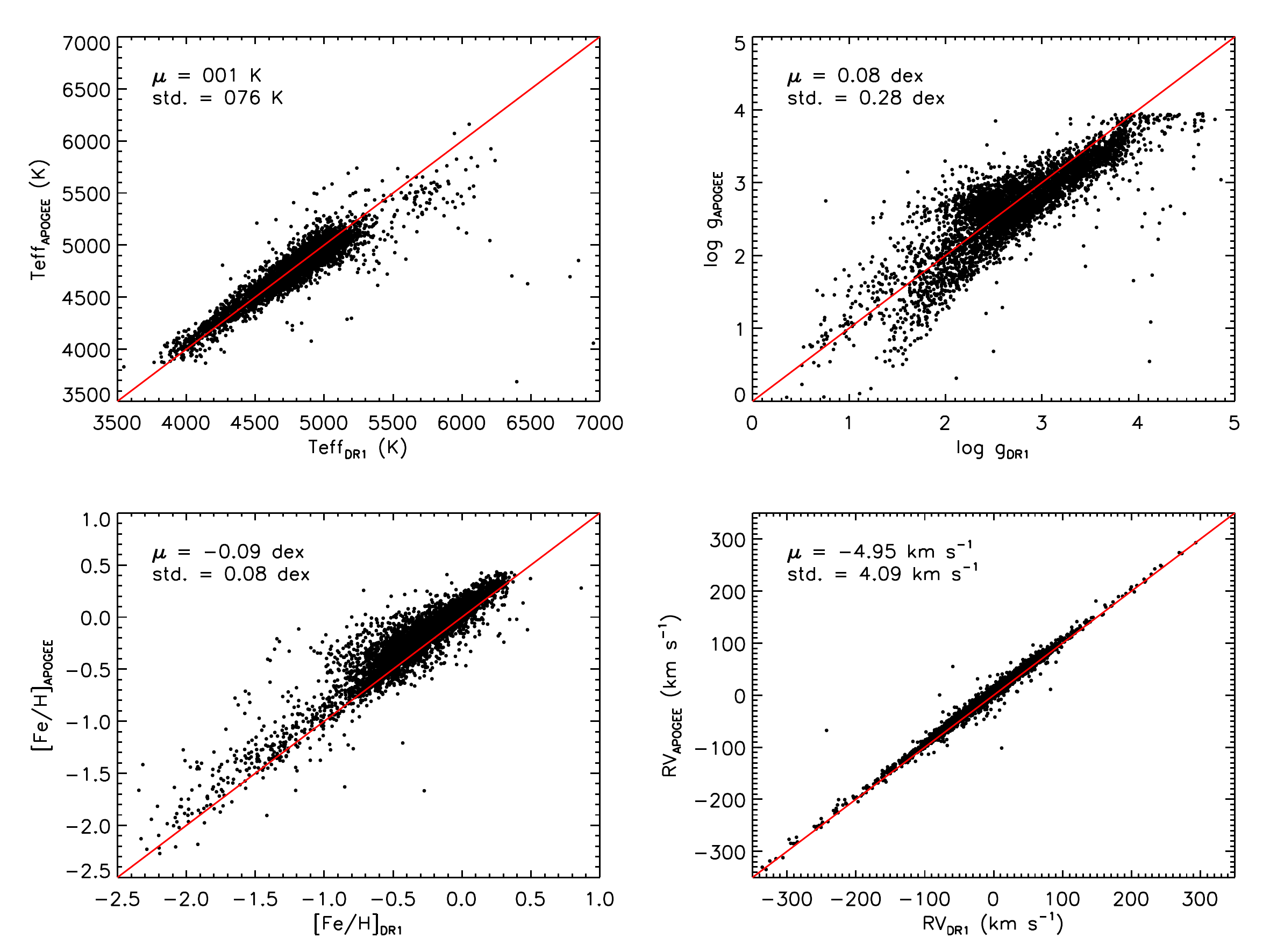}\\ 
\caption{{Comparison of stellar parameters for stars in common between LAMOST DR1 and APOGEE.}\label{figure11}}
\end{figure}

Zhang et al. (private communication) has also compared the parameters given by DR1 with those from APOGEE, as shown in Figure~\ref{figure11}. Besides these, \citet{2015inprep1} has compared 499 radial velocities (from DR1) with RVs of the same targets from MMT+Hectospec, which has RV accuracy of $\sim$2.5~km~s$^{-1}$ \citep{2009ApJ...703..441D,2012ApJ...750...97D}. The result is $V_r$(DR1)-$V_r$(MMT)=$-3.76\pm6.65$~km~s$^{-1}$ (see \citealt{2015inprep1}). 

\section{Data products}
LAMOST DR1 consists of 2,204,860 spectra of stars, galaxies, quasars, and other unknown type. The footprint on the sky is shown in Figure~\ref{figure12}, and the distributions of stellar parameters for all stars in RV vs. [Fe/H] and $\log{g}$ vs. $T_{\rm eff}$ are shown in Figure~\ref{figure13}. DR1 includes the following:

\begin{figure}[h]
\centering
\includegraphics[width=120mm]{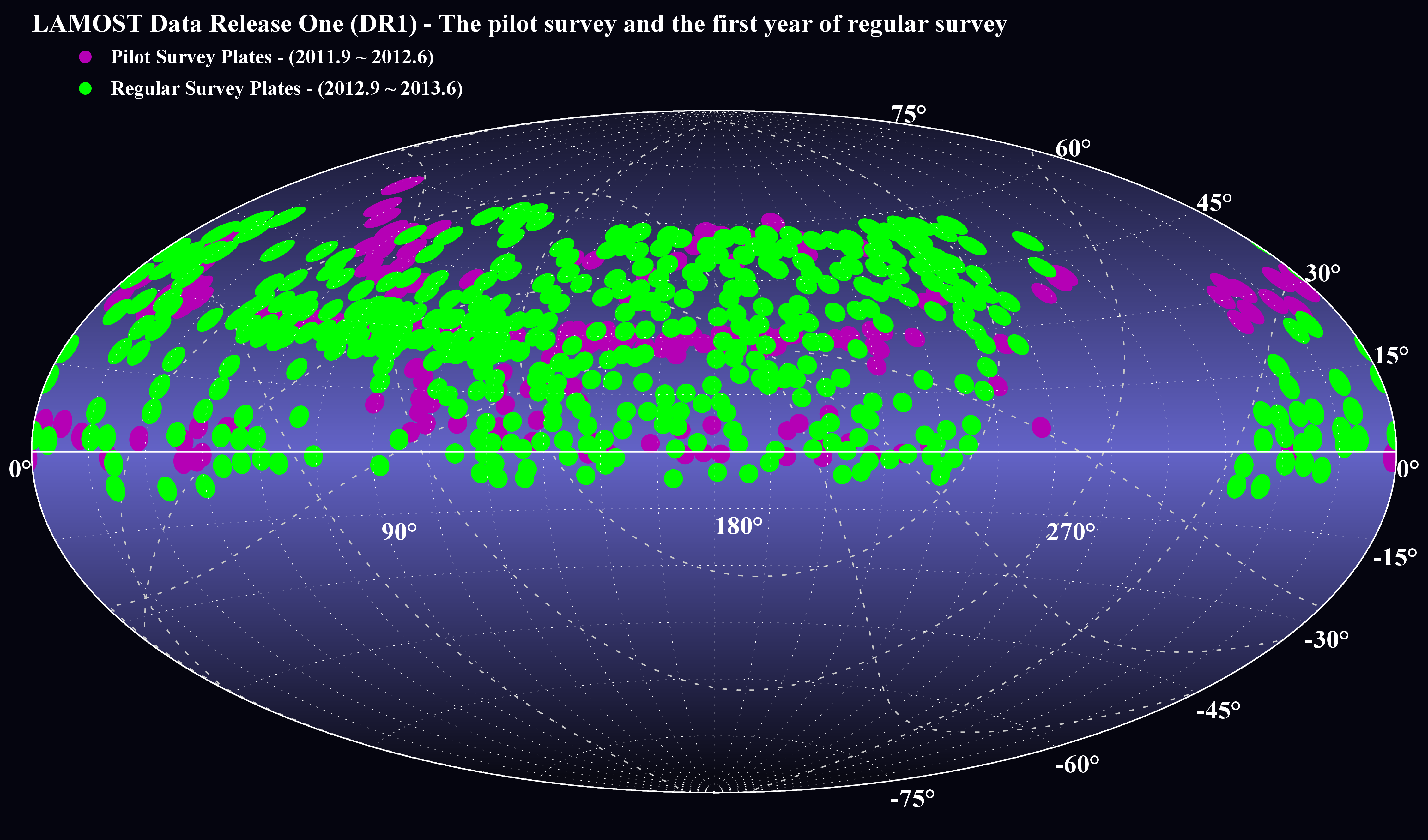}\\ 
\caption{{The footprints of LAMOST DR1 in equatorial candidates. Pilot survey plates are shown as purple circles, and the first year data are green points.}\label{figure12}}
\end{figure}

\begin{figure}[h]
\centering
\includegraphics[width=66mm]{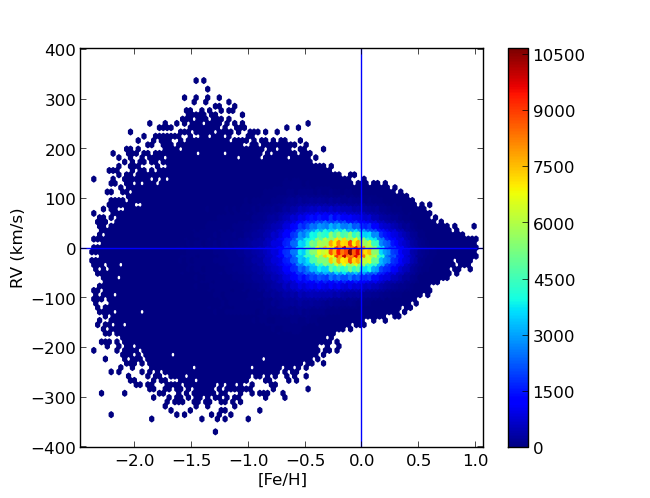}
\includegraphics[width=70mm]{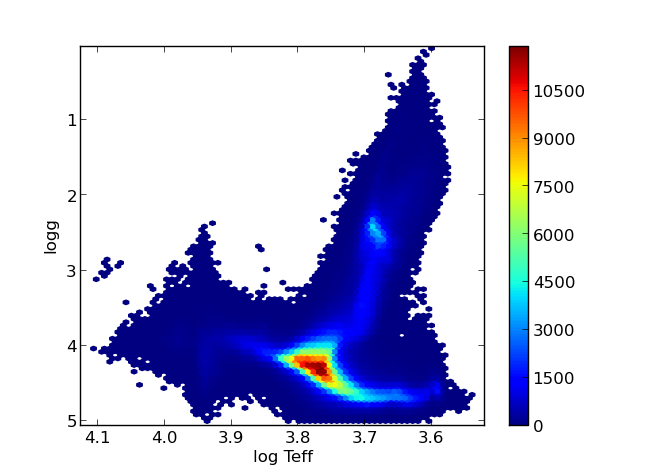}
\caption{{The distribution of stars in LAMOST DR1 in the planes of RV vs. [Fe/H] (left) and $\log{g}$ vs. $T_{\rm eff}$ (right).}\label{figure13}}
\end{figure}

\textit{1.	Spectra.} - In general, there are 2,204,860 flux- and wavelength-calibrated, sky-subtracted spectra released publicly in the DR1. It includes 1,944,329 stars, 12,082 galaxies, 5,017 quasars, and 243,355 unknown objects. These spectra cover the range 3690-9100~\AA~ with a resolution of 1800.

\textit{2.	Spectroscopic Parameters Catalogs.} - In this data release, four spectroscopic parameters catalogs are included: the DR1 general catalog, the A type stars catalog, the late A, F, G and K type stellar parameter catalog, and the M star catalog. All of these include 30 basic pieces of information about each spectrum; e.g., RA, Dec, SNR, magnitudes, spectral type, and redshift, among others. In addition, the A type stars catalog also publishes four Balmer line indices and four spectral line widths at 20\% below the local continuum, the late A, F, G and K type stars catalog provides effective temperature, surface gravity, and metallicity, and the M star catalog additionally contains a flag to indicate if a star shows magnetic activity.

\subsection{FITS File Description}
\subsubsection{FITS File Designation}
The file names of the released FITS files are in the form `spec-MMMMM-YYYY$\_$spXX-FFF.fits', where `MMMMM' represents the local modified Julian day (MJD), `YYYY' is the plan identity string (PLANID), `XX' indicates the spectrograph number (between 1 and 16), and `FFF' shows the fiber number (between 1 and 255). In addition, we provide a LAMOST designation for an object via the DESIG keyword that is formed as `LAMOST JHHMMSS.ss+DDMMSS.ss', where `HHMMSS.ss' is right ascension in units of HMS, and `+DDMMSS.ss' is declination in units of DMS.

\subsubsection{Primary FITS Header}
We reorganize keywords of the primary FITS header in DR1 and divided them into eight groups including mandatory keywords, file information keywords, telescope parameter keywords, observation parameter keywords, spectrograph parameters keywords, weather condition keywords, data reduction parameters keywords and spectra analysis results. See details in the appendix.

\subsubsection{Primary data array}
The primary data array has five rows and NAXIS1 (a keyword explained previously) columns. The five rows are Flux, Inverse Variance, Continuum-Subtracted Flux, Andmask and Ormask of each. It is noted that the `inverse variance' of the uncertainties (one over sigma-squared) can be used to estimate signal-to-noise ratio of each pixel ${(flux*(inverse~variance)^{0.5})}$, and `Andmask'/`Ormask' are mask flags of six quality situations for each pixel, respectively. The details of these two masks are listed in Table\ref{table11} in the appendix.

\subsection{LAMOST Catalogs}
Four LAMOST DR1 catalogs are released and available at the website \url{http://dr1.lamost.org/}, including the LAMOST general catalog, the A type stars catalog, the late A, F, G and K type stars catalog and the M star catalog. The detailed information is listed in tables (Table\ref{table13},Table\ref{table14},Table\ref{table15}) for each catalog.

The uncertainties of measurements in DR1 are not included since the error estimation is unreliable and needs to be improved; however, the uncertainties will be released in DR2.

\subsubsection{LAMOST general catalog}
The LAMOST general catalog includes not only 717,660 spectra from the LAMOST pilot survey, comprised of 648,820 stellar spectra, 2723 galaxy spectra, 621 quasar spectra, and 65496 unknown spectra, but also 1,487,200 spectra from the LAMOST general survey, which includes 1,295,586 stellar spectra, 9,359 galaxy spectra, 4,396 quasar spectra, and 177,859 unknown spectra. In this catalog, there are 1,221,538 spectra with SNR in the $g$-band larger than 10, 1,839,808 spectra with SNR in $i$-band larger than 10, and 1,198,410 spectra with SNR in both $g$- and $i$-bands larger than 10.
% and SNR of i band larger than 10.
 
All parameters of this catalog are listed in Table~\ref{table12}. For galaxies and quasars, `z' provides redshifts for them, and their `z' is set to -9999.00000000 if their redshift is unable to be estimated. For stars, `rv' is their heliocentric radial velocity, and the `rv' is set to -9999.00000000 if their rv is unable to be estimated. For objects with unknown type, their `z' and `rv' are both set to -9999.00000000. In addition, for a galaxy or a quasar, the value of `rv' is empty, and for a star, the value of `z' is null.

On the website \url{http://dr1.lamost.org}, we provide two format with the same contents for LAMOST general catalogs; one catalog is in .fits format, and the other is .csv format.

\subsubsection{A type stars catalog}
There are 100,073 A-type stellar spectra in this catalog both from the pilot and general survey. Table\ref{table13} shows all fields of this catalog. The `Class' field presents the two-dimensional spectral classification results, and nearly all A type stars have luminosity class provided by the LAMOST analysis pipeline. Balmer line indices are included in `Halpha$\_$Indice', `Hbeta$\_$Indice', `Hgama$\_$Indice', and `Hdelta$\_$Indice' fields, and `Halpha$\_$D0.2', `Hbeta$\_$D0.2', `Hgama$\_$D0.2', and 'Hdelta$\_$D0.2' are the widths at 20$\%$ below the local continuum of these four Balmer lines. However, these eight fields are temporarily empty, and will be released in September of this year. The .csv format A type stars catalog can be download from the website (\url{http://dr1.lamost.org}).

\subsubsection{AFGK  high quality stellar parameter catalog}
AFGK high quality stellar parameter catalog includes 61,686 A type stars, 548,214 F type stars, 253,275 G type stars and 222,229 K type stars. These spectra are selected with SNR$_g \geq 6$ from dark-night observations and SNR$_g \geq 20$ from bright nights. The effective temperatures ($T_{\rm eff}$), surface gravities ($\log{g}$) and metallicities (Fe/H) are determined by LASP. The A type stars in this catalog are those from the A type star catalog with high SNR spectra. These data are available from the website \url{http://dr1.lamost.org}.

\subsubsection{M star catalog}
The M star catalog contains 121,522 spectra, and their parameters are listed in Table\ref{table15}. The basic parameters are provided as other catalogs, particularly the magnetic activity flags `magact' are included in this catalog. The flag will be assigned to be 1 if stars have magnetic activity, 0 if no magnetic activity and -9999 if the magnetic activity could not be determined due to low SNR. The specific catalog for M dwarfs is given in \citep{Guo..RAA..inpre}. The .csv format M star catalog can be obtained from the website \url{http://dr1.lamost.org}. 

%%% stopped here

\subsection{Caveats}
\subsubsection{Radial velocity error}
The DR1 pipeline adopted local Beijing Time rather than UTC during the correction from geocentric coordinates to heliocentric coordinates. Thus an error in radical velocity is introduced, with mean value of 0.36~km~s$^{-1}$. In addition, the RV errors of some A type stars reach 80~km~s$^{-1}$ because their wavelength scale was incorrectly calibrated with air wavelengths instead of vacuum wavelengths. These are listed on the DR1 website.  

\subsubsection{Spectral line errors coming from sub--exposure combination}
The LAMOST spectra released in DR1 are combined from several sub--exposures. If one of the three sub--exposures is problematic, spectral lines may be erroneously scaled (clipped) during the B-spline fitting by using the wrong data points. There are 66,838 spectra affected in DR1, which are listed on DR1 website.

\subsubsection{Offset of very bright targets}
Some coordinates of bright objects are artificially shifted to avoid saturated exposures. However, we list the original coordinates in the DR1 website.

\subsubsection{Mistaken spectra due to fiber problem}
Problems were found for tens of thousands of spectra in DR1 that are due to fibers or positioners. These were discovered after the data release, and include several thousands of spectra that have stellar parameters $T_{\rm eff}$, $\log{g}$, and [Fe/H].  A separate list of these problematic spectra is provided on the DR1 website along with their IDs, and will be excluded in DR2.

\section{Data access and use}
\label{sect:data policy}
The LAMOST DR1 can be accessed at LAMOST Data Release web portal, \url{ http://dr1.lamost.org}, which is developed under the collaboration with Chinese Virtual Observatory (China-VO). Web form based database search and data retrieve are available for everyone. Advanced features are provided for registered users, for example batch search, bulk retrieve and personal space.

For registered users, lists of identifiers or coordinates supported by CDS SIMBAD\footnote{\url{http://simbad.u-strasbg.fr/simbad/}} and VizieR\footnote{\url{http://vizier.u-strasbg.fr/viz-bin/VizieR}}, or lists of LAMOST identifiers are acceptable to upload and use to search the database in batch mode. SQL-like queries are also supported. Virtual Observatory (VO) tools, such as Aladin\footnote{\url{http://aladin.u-strasbg.fr/aladin.gml}} and SpecView\footnote{\url{http://www.stsci.edu/institute/software_hardware/specview/}}, and tools developed specifically for LAMOST are integrated into the web portal. Search results and spectra can be displayed and analyzed on the fly. Large results and datasets can be retrieved in several ways, including FTP, wget, Cloud storage, and VO data access interfaces. 

Based on CDS X-Match Service\footnote{\url{http://cdsxmatch.u-strasbg.fr/xmatch}}, cross identification with dozens of well-known catalogs is also provided as an advanced feature of the data release portal.

According to the LAMOST data policy, any publication making use of LAMOST data should cite the standard acknowledgement as following 'Guoshoujing Telescope (the Large Sky Area Multi-Object Fiber Spectroscopic Telescope LAMOST) is a National Major Scientific Project built by the Chinese Academy of Sciences. Funding for the project has been provided by the National Development and Reform Commission. LAMOST is operated and managed by the National Astronomical Observatories, Chinese Academy of Sciences.'

\begin{acknowledgements}
The work was funded by the National Science Foundation of China (Grant Nos. 11390371) and the National Basic Research Program of China (973 Program, 2014CB845700). The Guo ShouJing Telescope (the Large Sky Area Multi-Object Fiber Spectroscopic Telescope, LAMOST) is a National Major Scientific Project built by the Chinese Academy of Sciences. Funding for the project has been provided by the National Development and Reform Commission. LAMOST is operated and managed by the National Astronomical Observatories, Chinese Academy of Sciences. The authors deeply thank Prof. Georges Comte for valuable discussions.
\end{acknowledgements}

\bibliography{reference}

\begin{thebibliography}{67}
\providecommand{\natexlab}[1]{#1}
\providecommand{\selectlanguage}[1]{\relax}

\bibitem[{Ai et~al.(2015)}]{Ai..RAA..inpre}
Ai, Y., et~al. 2015, inprep

\bibitem[{{Batalha} et~al.(2010){Batalha}, {Borucki}, {Koch}
  et~al.}]{2010ApJ...713L.109B}
{Batalha}, N.~M., {Borucki}, W.~J., {Koch}, D.~G., et~al. 2010, \apjl, 713,
  L109

\bibitem[{{Borucki} et~al.(2010){Borucki}, {Koch}, {Basri}
  et~al.}]{2010Sci...327..977B}
{Borucki}, W.~J., {Koch}, D., {Basri}, G., et~al. 2010, Science, 327, 977

\bibitem[{{Bovy} et~al.(2011){Bovy}, {Hennawi}, {Hogg}
  et~al.}]{2011ApJ...729..141B}
{Bovy}, J., {Hennawi}, J.~F., {Hogg}, D.~W., et~al. 2011, \apj, 729, 141

\bibitem[{{Bruntt} et~al.(2012){Bruntt}, {Basu}, {Smalley}
  et~al.}]{2012MNRAS.423..122B}
{Bruntt}, H., {Basu}, S., {Smalley}, B., et~al. 2012, \mnras, 423, 122

\bibitem[{{Carlin} et~al.(2013){Carlin}, {DeLaunay}, {Newberg}
  et~al.}]{2013ApJ...777L...5C}
{Carlin}, J.~L., {DeLaunay}, J., {Newberg}, H.~J., et~al. 2013, \apjl, 777, L5

\bibitem[{Cat et~al.(2015)}]{De..Unknown..inpre}
Cat, P.~D., et~al. 2015, inprep

\bibitem[{{Chen} et~al.(2014){Chen}, {Bai}, {Luo}, \&
  {Zhao}}]{2014SPIE.9149E..1NC}
{Chen}, J.-J., {Bai}, Z.-R., {Luo}, A.-L., \& {Zhao}, Y.-H. 2014, in Society of
  Photo-Optical Instrumentation Engineers (SPIE) Conference Series,
  \emph{Society of Photo-Optical Instrumentation Engineers (SPIE) Conference
  Series}, vol. 9149, 1

\bibitem[{{Chen} et~al.(2012){Chen}, {Hou}, {Yu} et~al.}]{2012RAA....12..805C}
{Chen}, L., {Hou}, J.-L., {Yu}, J.-C., et~al. 2012, Research in Astronomy and
  Astrophysics, 12, 805

\bibitem[{{Cui} et~al.(2012){Cui}, {Zhao}, {Chu} et~al.}]{2012RAA....12.1197C}
{Cui}, X.-Q., {Zhao}, Y.-H., {Chu}, Y.-Q., et~al. 2012, Research in Astronomy
  and Astrophysics, 12, 1197

\bibitem[{{Deng} et~al.(2012){Deng}, {Newberg}, {Liu}
  et~al.}]{2012RAA....12..735D}
{Deng}, L.-C., {Newberg}, H.~J., {Liu}, C., et~al. 2012, Research in Astronomy
  and Astrophysics, 12, 735

\bibitem[{{Dong} et~al.(2014){Dong}, {Zheng}, {Zhu}
  et~al.}]{2014ApJ...789L...3D}
{Dong}, S., {Zheng}, Z., {Zhu}, Z., et~al. 2014, \apjl, 789, L3

\bibitem[{{Drout} et~al.(2012){Drout}, {Massey}, \&
  {Meynet}}]{2012ApJ...750...97D}
{Drout}, M.~R., {Massey}, P., \& {Meynet}, G. 2012, \apj, 750, 97

\bibitem[{{Drout} et~al.(2009){Drout}, {Massey}, {Meynet}, {Tokarz}, \&
  {Caldwell}}]{2009ApJ...703..441D}
{Drout}, M.~R., {Massey}, P., {Meynet}, G., {Tokarz}, S., \& {Caldwell}, N.
  2009, \apj, 703, 441

\bibitem[{{Du} et~al.(2012){Du}, {Luo}, {Zhang}, {Wu}, \&
  {Wang}}]{2012SPIE.8451E..37D}
{Du}, B., {Luo}, A., {Zhang}, J., {Wu}, Y., \& {Wang}, F. 2012, in Society of
  Photo-Optical Instrumentation Engineers (SPIE) Conference Series,
  \emph{Society of Photo-Optical Instrumentation Engineers (SPIE) Conference
  Series}, vol. 8451, 37

\bibitem[{Gao et~al.(2015)Gao, Zhang, Xiang et~al.}]{Gao..RAA..inpre}
Gao, H., Zhang, H., Xiang, M., et~al. 2015, inprep

\bibitem[{Guo et~al.(2015)}]{Guo..RAA..inpre}
Guo, Y.-X., et~al. 2015, inprep

\bibitem[{{Huo} et~al.(2013){Huo}, {Liu}, {Xiang} et~al.}]{2013AJ....145..159H}
{Huo}, Z.-Y., {Liu}, X.-W., {Xiang}, M.-S., et~al. 2013, "AJ", 145, 159

\bibitem[{{Huo} et~al.(2010){Huo}, {Liu}, {Yuan} et~al.}]{2010RAA....10..612H}
{Huo}, Z.-Y., {Liu}, X.-W., {Yuan}, H.-B., et~al. 2010, Research in Astronomy
  and Astrophysics, 10, 612

\bibitem[{{Koleva} et~al.(2009){Koleva}, {Prugniel}, {Bouchard}, \&
  {Wu}}]{2009A&A...501.1269K}
{Koleva}, M., {Prugniel}, P., {Bouchard}, A., \& {Wu}, Y. 2009, \aap, 501, 1269

\bibitem[{{Lasker} et~al.(2008){Lasker}, {Lattanzi}, {McLean}
  et~al.}]{2008AJ....136..735L}
{Lasker}, B.~M., {Lattanzi}, M.~G., {McLean}, B.~J., et~al. 2008, \aj, 136, 735

\bibitem[{{Lee} et~al.(2008){Lee}, {Beers}, {Sivarani}
  et~al.}]{2008AJ....136.2022L}
{Lee}, Y.~S., {Beers}, T.~C., {Sivarani}, T., et~al. 2008, \aj, 136, 2022

\bibitem[{{Liu} et~al.(2014){Liu}, {Yuan}, {Huo} et~al.}]{2014IAUS..298..310L}
{Liu}, X.-W., {Yuan}, H.-B., {Huo}, Z.-Y., et~al. 2014, in IAU Symposium,
  \emph{IAU Symposium}, vol. 298, edited by S.~{Feltzing}, G.~{Zhao}, N.~A.
  {Walton}, \& P.~{Whitelock}, 310--321

\bibitem[{{Luo} et~al.(2012){Luo}, {Zhang}, {Zhao}
  et~al.}]{2012RAA....12.1243L}
{Luo}, A.-L., {Zhang}, H.-T., {Zhao}, Y.-H., et~al. 2012, Research in Astronomy
  and Astrophysics, 12, 1243

\bibitem[{Marquardt(1963)}]{marquardt1963algorithm}
Marquardt, D.~W. 1963, Journal of the Society for Industrial \& Applied
  Mathematics, 11, 431

\bibitem[{{Molenda-{\.Z}akowicz} et~al.(2013){Molenda-{\.Z}akowicz}, {Sousa},
  {Frasca} et~al.}]{2013MNRAS.434.1422M}
{Molenda-{\.Z}akowicz}, J., {Sousa}, S.~G., {Frasca}, A., et~al. 2013, \mnras,
  434, 1422

\bibitem[{Mor{\'e} et~al.(1980)Mor{\'e}, Garbow, \& Hillstrom}]{more1980user}
Mor{\'e}, J.~J., Garbow, B.~S., \& Hillstrom, K.~E. 1980, User guide for
  MINPACK-1, Tech. rep., CM-P00068642

\bibitem[{{Peng} et~al.(2012){Peng}, {Zhang}, {Zhao}, \&
  {Wu}}]{2012MNRAS.425.2599P}
{Peng}, N., {Zhang}, Y., {Zhao}, Y., \& {Wu}, X.-b. 2012, \mnras, 425, 2599

\bibitem[{{Perryman} et~al.(1997){Perryman}, {Lindegren}, {Kovalevsky}
  et~al.}]{1997A&A...323L..49P}
{Perryman}, M.~A.~C., {Lindegren}, L., {Kovalevsky}, J., et~al. 1997, \aap,
  323, L49

\bibitem[{{Prugniel} \& {Soubiran}(2001)}]{2001A&A...369.1048P}
{Prugniel}, P., \& {Soubiran}, C. 2001, \aap, 369, 1048

\bibitem[{{Prugniel} \& {Soubiran}(2004)}]{2004astro.ph..9214P}
{Prugniel}, P., \& {Soubiran}, C. 2004, ArXiv Astrophysics e-prints

\bibitem[{{Prugniel} et~al.(2007){Prugniel}, {Soubiran}, {Koleva}, \& {Le
  Borgne}}]{2007astro.ph..3658P}
{Prugniel}, P., {Soubiran}, C., {Koleva}, M., \& {Le Borgne}, D. 2007, ArXiv
  Astrophysics e-prints

\bibitem[{{Ren} et~al.(2013){Ren}, {Luo}, {Li} et~al.}]{2013AJ....146...82R}
{Ren}, J., {Luo}, A., {Li}, Y., et~al. 2013, \aj, 146, 82

\bibitem[{{Ren} et~al.(2007){Ren}, {Wang}, {Li} et~al.}]{2007AcASn..48..500R}
{Ren}, J., {Wang}, J., {Li}, F.~H., K., et~al. 2007, Acta Astronomica Sinica,
  48, 500

\bibitem[{{Roeser} et~al.(2010){Roeser}, {Demleitner}, \&
  {Schilbach}}]{2010AJ....139.2440R}
{Roeser}, S., {Demleitner}, M., \& {Schilbach}, E. 2010, \aj, 139, 2440

\bibitem[{{Schneider} et~al.(2010){Schneider}, {Richards}, {Hall}
  et~al.}]{2010yCat.7260....0S}
{Schneider}, D.~P., {Richards}, G.~T., {Hall}, P.~B., et~al. 2010, VizieR
  Online Data Catalog, 7260, 0

\bibitem[{{Shi} et~al.(2014{\natexlab{a}}){Shi}, {Comte}, {Luo}
  et~al.}]{2014A&A...564A..89S}
{Shi}, Z.~X., {Comte}, G., {Luo}, A.~L., et~al. 2014{\natexlab{a}}, \aap, 564,
  A89

\bibitem[{{Shi} et~al.(2014{\natexlab{b}}){Shi}, {Luo}, {Comte}
  et~al.}]{2014RAA....14.1234S}
{Shi}, Z.-X., {Luo}, A.-L., {Comte}, G., et~al. 2014{\natexlab{b}}, Research in
  Astronomy and Astrophysics, 14, 1234

\bibitem[{{Song} et~al.(2012){Song}, {Luo}, {Comte}
  et~al.}]{2012RAA....12..453S}
{Song}, Y.-H., {Luo}, A.-L., {Comte}, G., et~al. 2012, Research in Astronomy
  and Astrophysics, 12, 453

\bibitem[{{Stoughton} et~al.(2002){Stoughton}, {Lupton}, {Bernardi}, \&
  {Blanton}}]{2002AJ....123..485S}
{Stoughton}, C., {Lupton}, R.~H., {Bernardi}, M., \& {Blanton}, M.~R. 2002,
  \aj, 123, 485

\bibitem[{{Thygesen} et~al.(2012){Thygesen}, {Frandsen}, {Bruntt}
  et~al.}]{2012A&A...543A.160T}
{Thygesen}, A.~O., {Frandsen}, S., {Bruntt}, H., et~al. 2012, \aap, 543, A160

\bibitem[{{Wu} et~al.(2010{\natexlab{a}}){Wu}, {Chen}, {Jia}
  et~al.}]{2010RAA....10..737W}
{Wu}, X.-B., {Chen}, Z.-Y., {Jia}, Z.-D., et~al. 2010{\natexlab{a}}, Research
  in Astronomy and Astrophysics, 10, 737

\bibitem[{{Wu} et~al.(2012){Wu}, {Hao}, {Jia}, {Zhang}, \&
  {Peng}}]{2012AJ....144...49W}
{Wu}, X.-B., {Hao}, G., {Jia}, Z., {Zhang}, Y., \& {Peng}, N. 2012, \aj, 144,
  49

\bibitem[{{Wu} \& {Jia}(2010)}]{2010MNRAS.406.1583W}
{Wu}, X.-B., \& {Jia}, Z. 2010, \mnras, 406, 1583

\bibitem[{{Wu} et~al.(2010{\natexlab{b}}){Wu}, {Jia}, {Chen}
  et~al.}]{2010RAA....10..745W}
{Wu}, X.-B., {Jia}, Z.-D., {Chen}, Z.-Y., et~al. 2010{\natexlab{b}}, Research
  in Astronomy and Astrophysics, 10, 745

\bibitem[{{Wu} et~al.(2014{\natexlab{a}}){Wu}, {Luo}, {Du}, \&
  {Guo}}]{2014IAUS..298..445W}
{Wu}, Y., {Luo}, A., {Du}, B., \& {Guo}, Y. 2014{\natexlab{a}}, in IAU
  Symposium, \emph{IAU Symposium}, vol. 298, edited by S.~{Feltzing},
  G.~{Zhao}, N.~A. {Walton}, \& P.~{Whitelock}, 445--445

\bibitem[{{Wu} et~al.(2014{\natexlab{b}}){Wu}, {Luo}, {Du}, {Zhao}, \&
  {Yuan}}]{2014arXiv1407.1980W}
{Wu}, Y., {Luo}, A., {Du}, B., {Zhao}, Y., \& {Yuan}, H. 2014{\natexlab{b}},
  ArXiv e-prints

\bibitem[{{Wu} et~al.(2011{\natexlab{a}}){Wu}, {Luo}, {Li}
  et~al.}]{2011RAA....11..924W}
{Wu}, Y., {Luo}, A.-L., {Li}, H.-N., et~al. 2011{\natexlab{a}}, Research in
  Astronomy and Astrophysics, 11, 924

\bibitem[{{Wu} et~al.(2011{\natexlab{b}}){Wu}, {Singh}, {Prugniel}, {Gupta}, \&
  {Koleva}}]{2011A&A...525A..71W}
{Wu}, Y., {Singh}, H.~P., {Prugniel}, P., {Gupta}, R., \& {Koleva}, M.
  2011{\natexlab{b}}, \aap, 525, A71

\bibitem[{{Xiang} et~al.(2015){Xiang}, {Liu}, {Yuan}
  et~al.}]{2015MNRAS.448..822X}
{Xiang}, M.~S., {Liu}, X.~W., {Yuan}, H.~B., et~al. 2015, \mnras, 448, 822

\bibitem[{{Yang} et~al.(2012){Yang}, {Carlin}, {Liu}
  et~al.}]{2012RAA....12..781Y}
{Yang}, F., {Carlin}, J.~L., {Liu}, C., et~al. 2012, Research in Astronomy and
  Astrophysics, 12, 781

\bibitem[{Yang et~al.(2015)}]{Yang..MNRAS..accepted}
Yang, M., et~al. 2015, accepted

\bibitem[{{Yanny} et~al.(2009){Yanny}, {Rockosi}, {Newberg}
  et~al.}]{2009AJ....137.4377Y}
{Yanny}, B., {Rockosi}, C., {Newberg}, H.~J., et~al. 2009, \aj, 137, 4377

\bibitem[{{Yao} et~al.(2012){Yao}, {Liu}, {Zhang} et~al.}]{2012RAA....12..772Y}
{Yao}, S., {Liu}, C., {Zhang}, H.-T., et~al. 2012, Research in Astronomy and
  Astrophysics, 12, 772

\bibitem[{{Yi} et~al.(2014){Yi}, {Luo}, {Song} et~al.}]{2014AJ....147...33Y}
{Yi}, Z., {Luo}, A., {Song}, Y., et~al. 2014, \aj, 147, 33

\bibitem[{{York} et~al.(2000){York}, {Adelman}, {Anderson}
  et~al.}]{2000AJ....120.1579Y}
{York}, D.~G., {Adelman}, J., {Anderson}, J.~E., Jr., et~al. 2000, \aj, 120,
  1579

\bibitem[{{Yuan} et~al.(2014{\natexlab{a}}){Yuan}, {Liu}, {Huo}
  et~al.}]{2014arXiv1412.6628Y}
{Yuan}, H., {Liu}, X., {Huo}, Z., et~al. 2014{\natexlab{a}}, ArXiv e-prints

\bibitem[{{Yuan} et~al.(2012){Yuan}, {Zhang}, {Lei}, \&
  {Dong}}]{2012SPIE.8448E..2AY}
{Yuan}, H., {Zhang}, H., {Lei}, Y., \& {Dong}, Y. 2012, in Society of
  Photo-Optical Instrumentation Engineers (SPIE) Conference Series,
  \emph{Society of Photo-Optical Instrumentation Engineers (SPIE) Conference
  Series}, vol. 8448, 2

\bibitem[{{Yuan} et~al.(2014{\natexlab{b}}){Yuan}, {Zhang}, {Zhang}, {Lei}, \&
  {Dong}}]{2014IAUS..298..452Y}
{Yuan}, H., {Zhang}, H., {Zhang}, Y., {Lei}, Y., \& {Dong}, Y.
  2014{\natexlab{b}}, in IAU Symposium, \emph{IAU Symposium}, vol. 298, edited
  by S.~{Feltzing}, G.~{Zhao}, N.~A. {Walton}, \& P.~{Whitelock}, 452--452

\bibitem[{{Yuan} et~al.(2010){Yuan}, {Liu}, {Huo} et~al.}]{2010RAA....10..599Y}
{Yuan}, H.-B., {Liu}, X.-W., {Huo}, Z.-Y., et~al. 2010, Research in Astronomy
  and Astrophysics, 10, 599

\bibitem[{{Zhang} et~al.(2015){Zhang}, {Huang}, \& {Liu}}]{2015inprep1}
{Zhang}, H., {Huang}, Y., \& {Liu}, X. 2015, inprep

\bibitem[{{Zhang} \& {Zhao}(2003)}]{2003PASP..115.1006Z}
{Zhang}, Y., \& {Zhao}, Y. 2003, \pasp, 115, 1006

\bibitem[{{Zhang} et~al.(2012){Zhang}, {Carlin}, {Yang}
  et~al.}]{2012RAA....12..792Z}
{Zhang}, Y.-Y., {Carlin}, J.~L., {Yang}, F., et~al. 2012, Research in Astronomy
  and Astrophysics, 12, 792

\bibitem[{{Zhang} et~al.(2013){Zhang}, {Deng}, {Liu}
  et~al.}]{2013AJ....146...34Z}
{Zhang}, Y.-Y., {Deng}, L.-C., {Liu}, C., et~al. 2013, \aj, 146, 34

\bibitem[{{Zhao} et~al.(2012){Zhao}, {Zhao}, {Chu}, {Jing}, \&
  {Deng}}]{2012RAA....12..723Z}
{Zhao}, G., {Zhao}, Y.-H., {Chu}, Y.-Q., {Jing}, Y.-P., \& {Deng}, L.-C. 2012,
  Research in Astronomy and Astrophysics, 12, 723

\bibitem[{{Zhao} et~al.(2013){Zhao}, {Luo}, {Oswalt}, \&
  {Zhao}}]{2013AJ....145..169Z}
{Zhao}, J.~K., {Luo}, A.~L., {Oswalt}, T.~D., \& {Zhao}, G. 2013, \aj, 145, 169

\bibitem[{{Zou} et~al.(2011){Zou}, {Yang}, {Zhang}
  et~al.}]{2011RAA....11.1093Z}
{Zou}, H., {Yang}, Y.-B., {Zhang}, T.-M., et~al. 2011, Research in Astronomy
  and Astrophysics, 11, 1093

\end{thebibliography}
\newpage
\appendix
\section*{Appendix : Table List}
There are nine tables for FITS files, including: Mandatory keywords(Table\ref{table3}), File Information Keywords(Table\ref{table4}), Telescope Parameter Keywords(Table\ref{table5}), Observation Parameter Keywords(Table\ref{table6}), Spectrograph Parameters Keywords(Table\ref{table7}), Weather Condition Keywords(Table\ref{table8}), Data Reduction Parameters Keywords(Table\ref{table9}), Spectra Analysis Results Keyword(Table\ref{table10}) and the significance of six bits of `Andmask' and `Ormask'(Table\ref{table11}). These values must be written in fixed format.

There are another four tables, the gerneral catalog(Table\ref{table12}), A-type star catalog(Table\ref{table13}), Late A F G K catalog(Table\ref{table14}),  and M-type star catalog(Table\ref{table15}).   \\

%%%%%%%%%%%%%%%%%table3
\begin{table}[h]
\begin{center}
\caption{{Mandatory Keywords}\label{table3}}
\begin{tabular}[]{lllll}
\hline
Keywords 		&&	Value		&& 	Comment\\
\hline
SIMPLE  		&&	T 		&&	Primary Header created by MWRFITS v1.8    \\
BITPIX  		&&    -32 		&&									\\
NAXIS   		&&    2 		&&									\\
NAXIS1  		&&    3909 		&&									\\
NAXIS2  		&&    5 		&&									\\
EXTEND  		&&    T 		&&	Extensions may be present				\\
\hline
\end{tabular}
\begin{flushleft}
{\sc Notes:}\\
\textbf{1. SIMPLE} It is required to be the first keyword in the primary header of all FITS file. The value field shall contain a logical constant with the value T if the file conforms to this standard. This keyword is mandatory for the primary header and is not permitted in extension headers. A value of F signifies that the file does not conform to this standard.\\
\textbf{2. BITPIX} The value field shall contain an integer, and it shall specify the number of bits that represent a data value. A value of -32 represents IEEE single precision floating point.\\
\textbf{3. NAXIS}The value field shall contain a non-negative integer no greater than 999, representing the number of axes in the associated data array. A value of zero signifies that no data follow the header in the HDU.\\
\textbf{4. NAXIS1 and NAXIS2}The value field of these two indexed keywords shall contain a non-negative integer, representing the number of elements along axis n of a data array. The NAXIS1 keyword represents the number of wavelength array, i.e., the column number of the primary data array, and the NAXIS2 keyword indicates the row number of the primary data array.\\
\textbf{5. EXTEND}The value field shall contain a logical value indicating whether the FITS file is allowed to contain conforming extensions following the primary HDU. This keyword may only appear in the primary header and must not appear in an extension header. If the value field is T then there may be conforming extensions in the FITS file following the primary HDU. This keyword is only advisory, so its presence with a value T does not require that the FITS file contains extensions, nor does the absence of this keyword necessarily imply that the file does not contain extensions[6].\\
\end{flushleft}
\end{center}
\end{table}

%%%%%%%%%%%%%%%%%%%table4
\begin{table}
\begin{center}
\caption{{File Information Keywords}\label{table4}}
\begin{tabular}[]{lllll}
\hline
Keywords 		&	Value		& 	Comment\\
\hline
FILENAME		& `spec-55859-F5902$\_$sp01-001.fits'	 &	\\
AUTHOR  		& `LAMOST Pipeline'   					 & Who compiled the information					\\
N$\_$EXTEN		&  1  									 & The extension number							\\
EXTEN0  		& `Flux Inverse Subcontinuum Andmask Ormask'  		 &		\\
ORIGIN  		& `NAOC-LAMOST'        						 & Organization \\
DATE    		& `2013-07-17T10:23:12' 						 & (UTC)				\\
\hline
\end{tabular}
\begin{flushleft}
{\sc Notes:}\\
\textbf{1. FILENAME}  The value field shall contain a character string giving the name of this FITS file. Take the `spec-55859-F5902$\_$sp01-001.fits' as an example, `55859’ is the local modified Julian day, `F5902’ is the plan ID, `sp01’ is the spectrograph ID, and `001’ is the Fiber ID.\\
\textbf{2. AUTHOR} This keyword contains a string constant ‘LAMOST Pipline’, which represents the author who produce this file.\\
\textbf{3. N$\_$EXTEN} The value field shall contain an integer giving the extension number of a FITS file.\\
\textbf{4. EXTEN0} This keyword contains a string constant ‘Flux Inverse Subcontinuum Andmask Ormask’ explaining each row of the primary data array in a primary HDU.\\
\textbf{5. ORIGIN} This ORIGIN keyword contains a string constant ‘NAOC-LAMOST’, which indicates the Organization responsible for this FITS file. ‘NAOC’ represents the abbreviation of National Astronomical Observatories, Chinese Academy of Sciences.\\
\textbf{6. DATE}The value field shall contain a character string giving the UTC time when this FITS file is created.\\
\end{flushleft}
\end{center}
\end{table}

%%%%%%%%%%%%%%%%%%%%%table5
\begin{table}
\begin{center}
\caption{{Telescope Parameter Keywords}\label{table5}}
\begin{tabular}[]{lllllllll}
\hline
Keywords 		&&&&	Value			&&&& 	Comment				\\
\hline
TELESCOP		&&&&	 `LAMOST'		&&&&   GuoShouJing Telescope		\\
LONGITUD		&&&&     117.580 		&&&&  [degree] Longitude of site	\\
LATITUDE		&&&&     40.3900 		&&&&  [degree] Latitude of site	\\
FOCUS   		&&&&     19964 		&&&&  [mm] Telescope focus		\\
CAMPRO  		&&&&	 `NEWCAM'       &&&&  Camera program name		\\
CAMVER  		&&&&	 `v2.0 '		&&&&  Camera program version		\\
\hline
\end{tabular}
\begin{flushleft}
{\sc Notes:}\\
\textbf{1. TELESCOP} This keyword contain a string constant `LAMOST' giving the name of our telescope.\\
\textbf{2. LONGITUD} The keyword contain a floating-point constant, which provide the longitude of Xinglong station where LAMOST is mounted on.\\
\textbf{3. LATITUDE} The keyword contain a floating-point constant, which provide the latitude of Xinglong station.\\
\textbf{4.	FOCUS} The keyword gives the telescope focus, and its unit is millimeter.\\
\textbf{5. CAMPRO} The value field contain a string constant `NEWCAM', which shows the name of camera.\\ 
\textbf{6. CAMVER} The value field contain a character string `v2.0', which gives the present camera program version.\\
\end{flushleft}
\end{center}
\end{table}

%%%%%%%%%%%%%%%%%%%%%table6
\begin{table}
\begin{center}
\caption{{Observation Parameter Keywords}\label{table6}}
\begin{tabular}[]{lllll}
\hline
Keywords 		&	Value			& 	Comment				\\
\hline
DATE-OBS		&	 `2013-06-03T16:47:08.38'  & The observation median UTC	\\
DATE-BEG		&	 `2013-06-04T00:13:52.0' 	& The observation start local time\\
DATE-END		&	`2013-06-04T01:21:17.0' 	& The observation end local time\\
MJD     		&     56447 				& Local Modified Julian Day	\\
MJDLIST 		&	 `56447-56447'        	& Local Modified Julian Day list\\
MJMLIST 		&	 `81283693-81283731'  	& Local Modified Julian Minute list 		\\
PLANID  		&	 `HD174640N254456F01' 	& Plan ID in use			\\
RA      		&	 266.506521500 			& [degree] Right ascension of object		\\
DEC     		&     23.7254148000 			& [degree] Declination of object		\\
DESIG   		&	 `LAMOST J174601.56+234331.4' & Designation of LAMOST target		\\
FIBERID 		&     80 					& Fiber ID of Object					\\
CELL$\_$ID 	&	 `E0818   '           	& Fiber Unit ID on the focal plane		\\
X$\_$VALUE 	&     51.4330635340 			& [mm] X coordinate of object on the focal plane 	\\
Y$\_$VALUE 	&    703.039784897 			& [mm] Y coordinate of object on the focal plane	\\
OBJNAME 		& 	`230388178985208'    		& Name of object				\\
OBJTYPE 		&	 `Star'           		& Object type from input catalog	\\
OBJSOURC		&	 `LEGUE$\_$LCH'          	& Name of input catalog			\\
FIBERTYP		&	 `Obj'           		& Fiber type of object			\\
MAGTYPE 		&	 `gri'           		& Magnitude type of object			\\
MAG1    		&     17.94 				& [magnitude] Mag1 of object		\\
MAG2    		&     17.47 				& [magnitude] Mag2 of object		\\
MAG3    		&     17.24 				& [magnitude] Mag3 of object		\\
MAG4    		&     99.00 				& [magnitude] Mag4 of object		\\
MAG5    		&     99.00 				& [magnitude] Mag5 of object		\\
MAG6    		&     99.00 				& [magnitude] Mag6 of object		\\
MAG7    		&     99.00 				& [magnitude] Mag7 of object		\\
OBS$\_$TYPE	&	 `OBJ'           		& The type of target (OBJ, FLAT, ARC or BIAS)\\	
OBSCOMM 		&	 `Science ' 			& Science, Test					\\
RADECSYS		&	 `FK5'           		& Equatorial coordinate system		\\
EQUINOX 		&     2000.00 				& Equinox in years				\\
\hline
\end{tabular}
\begin{flushleft}
\tiny
{\sc Notes:}\\
\textbf{1. DATE-OBS} The value field shall contain a character string, which gives the median moment UTC of multiple exposures.\\
\textbf{2. DATE-BEG} The value field shall contain a character string giving the observation start Beijing Time.\\
\textbf{3. DATE-END} The value field shall contain a character string, which provide the observation end Beijing Time.\\
\textbf{4. MJD} The value field shall a non-negative integer giving the local modified Julian day.\\
\textbf{5.	MJDLIST} The value field shall contain a character string, which shows a list of local modified Julian day of n times exposures.\\
\textbf{6. MJMLIST} The value field shall contain a character string, which shows a list of local modified Julian minute of n times exposures.\\
\textbf{7. PLANID} The value field shall contain a character string providing the plan ID of the target.\\
\textbf{8.	RA} The value field shall contain a non-negative real floating-point number, which gives the right ascension of target.\\
\textbf{9.	DEC} The value field shall contain a non-negative real floating-point number, which gives the declination of target.\\
\textbf{10. DESIG} The value field shall contain a character string, which indicates the name of LAMOST target. Like the name of SDSS target, numbers after the character ‘J’ and before ‘+’ represents RA in unit of HMS , and numbers after the character ‘+’ are DEC in unit of DMS.\\
\textbf{11. FIBERID} The value field shall contain a non-negative integer between 1 and 250, which shows the fiber ID and shall be used together with the spectrograph ID.\\
\textbf{12. CELL$\_$ID}  The value field shall contain a character string, which gives the fiber unit ID on the focal plane. LAMOST focal plane is divided into four quadrant named ‘EFGH’ respectively, the first character of this keyword represents the quadrant number, the first two numbers after the first character is the row number in this quadrant, and the next two numbers is the column numbers.\\
\textbf{13. X$\_$VALUE and Y$\_$VALUE} Their value field shall contain two real floating-point numbers, which give X and Y coordinates of target on the focal plane.\\
\textbf{14. OBJNAME} The value field shall contain character string, giving the name ID of object that determined by the RA, DEC and HTM method[7].\\
\textbf{15. OBJTYPE} The value field shall contain a character string giving the class of objects in input catalogs.\\ 
\textbf{16. OBJSOURC} The value field shall contain a character string which shows the name of organization or person who submit input catalog.\\
\textbf{17. FIBERTYP} The value field shall contain a character string, giving the type of fiber assigned to this target. This keyword has five values, i.e., Obj, Sky, F-std, Unused, PosErr and Dead. Obj means the fiber is assigned to a object, including star, galaxy and so on. Sky indicates the fiber is allocated to take skylight. F-std shows the fiber is used to take the light of a flux calibration standard star. Unused, PosErr and Dead mean the fiber is not used, goes to a wrong position, or does not work respectively.\\
\textbf{18. MAGTYPE} The value field shall contain a character string, which shows the magnitude type of a target.\\
\textbf{19. MAG1, MAG2, MAG3, MAG4, MAG5, MAG6 and MAG7} The value field shall contain a real floating-point number between 0 and 100, giving the associated magnitudes of MAGTYPE keyword. For example, The MAGTYPE keyword is ‘ugrizjh’, the MAG1, MAG2, MAG3, MAG4, MAG5, MAG6 and MAG7 keywords provide the magnitudes of u, g, r, i, z, j and h filter respectively.\\
\textbf{20. OBS$\_$TYPE} The value field shall contain a character string giving the type of observation targets, which include object, flat, bias and arc lamp.\\
\textbf{21. OBSCOMM}The value field shall contain a character string constant representing the observation purposes , which includes observations used for science researches and kinds of tests.\\
\textbf{22. RADECSYS}The value field shall contain a character string giving the equatorial coordinate system based on the J2000 position.\\
\textbf{23. EQUINOX} The value field shall contain a real floating-point number giving the standard epoch used at present.\\
\end{flushleft}
\end{center}
\end{table}

%%%%%%%%%%%%%%%%%%%%%table7
\begin{table}
\begin{center}
\caption{{Spectrograph Parameters Keywords}\label{table7}}
\begin{tabular}[]{lllll}
\hline
Keywords 		&&	Value			&& 	Comment				\\
\hline
SPID   		&&    1			&&	 Spectrograph ID				\\
SPRA   		&&    266.640 		&&	 [degree] Average RA of this spectrograph	\\
SPDEC   		&&    23.7144 		&&	 [degree] Average DEC of this spectrograph	\\
SLIT$\_$MOD	&&	 `x2/3 '        &&     Slit mode, x1 ,x2/3 or x1/2			\\
\hline
\end{tabular}
\begin{flushleft}
{\sc Notes:}\\
\textbf{1. SPID} The value field shall contain a non-integer numbers between 1 and 16, which provides the spectrograph ID.\\
\textbf{2. SPRA and SPDEC} The value field of these two keywords shall contain two real floating-point numbers, which are the averages of RA and DEC of all objects in each spectrograph.\\
\textbf{3. SLIT$\_$MOD} The value field shall contain a character string giving the mode of slit, which includes x1, x2/3 and x1/2. At present, only mode x2/3 is available, which responds spectra resolution equals to 1800.
\end{flushleft}
\end{center}
\end{table}

%%%%%%%%%%%%%%%%%%%%%%%%%table8
\begin{table}
\begin{center}
\caption{{Weather Condition Keywords}\label{table8}}
\begin{tabular}[]{lllll}
\hline
Keywords 		&&	Value			&& 	Comment				\\
\hline
TEMPCCDB		&&    -101.60 		&& 	[C degree] The temperature of blue CCD	\\
TEMPCCDR		&&    -103.90 		&&	[C degree] The temperature of red CCD	 	\\
SEEING  		&&    3.60 		&&	[arcsecond] Seeing during exposure		\\
MOONPHA 		&&    24.51		&& 	[day] Moon phase for a 29.53 days period	\\
TEMP$\_$AIR	&&    20.35 		&&	[C degree] Temperature outside dome		\\
DEWPOINT		&&    3.11 		&&	[C degree]							\\
DUST    		&&    '        '     &&	Reservation						\\
HUMIDITY		&&	0.32 			&&									\\
WINDD   		&&   87.00 		&& 	[degree] Wind direction					\\
WINDS   		&&   2.30 			&&	[m/s] Wind speed						\\
\hline
\end{tabular}
\begin{flushleft}
{\sc Notes:}\\
\textbf{1. TEMPCCDB} The value field shall contain a real floating-point number, which provides the temperature of blue CCD. The unit ‘C degree’ represents centigrade degree.\\
\textbf{2. TEMPCCDR} The value field shall contain a real floating-point number, which provides the temperature of red CCD. The unit ‘C degree’ represents centigrade degree.\\
\textbf{3. SEEING}  The value field shall contain a real floating-point number giving seeing during exposure, which is calculated by manually measuring the full width at half maximum of guide star image.\\
\textbf{4. MOONPHA} The value field shall contain a real floating-point number giving the moon phase.\\
\textbf{5. TEMP$\_$AIR} The value field shall contain a real floating-point number giving the temperature outside dome, which is measured by automatic weather instrument. The unit ‘C degree’ represents centigrade degree.\\
\textbf{6. DEWPOINT} The value field shall contain a real floating-point number giving the dew-point temperature, which is also measured by the automatic weather instrument. The unit ‘C degree’ represents centigrade degree.\\
\textbf{7. DUST} The value of this keyword is temporarily empty at present, because the dust measuring instrument is now in debugging, and we will write this parameters into fits header when problems are resolved.\\
\textbf{8. HUMIDITY} The value field shall contain a real floating-point number between 0 and 1, which gives humidity in the air.\\
\textbf{9. WINDD} The value field shall contain a real floating-point number which records the instantaneous wind direction when start exposure, and the direction of north is the 0 degree wind direction.\\
\textbf{10. WINDS} The value field shall contain a real floating-point number which records the instantaneous wind speed when start exposure, and wind direction and speed are also measured also by the automatic weather instrument.\\
\end{flushleft}
\end{center}
\end{table}

%%%%%%%%%%%%%%%%%%%%%%%%%table9
\begin{table}
\begin{center}
\caption{{Data Reduction Parameters Keywords(1)}\label{table9}}
\begin{tabular}[]{lllll}
\hline
Keywords 		&	Value						& 	Comment				\\
\hline
LAMPLIST		& `lamphgcdne.dat'   				& 	Arc lamp emission line list  		\\
SKYLIST 		& `skylines.dat'     				&	Sky emission line list 			\\
EXPID01 		& `01b-20130604001350-4-81283693' 	& 	ID string for exposure 1 	\\
EXPID02 		& `01b-20130604005112-5-81283731' 	&	ID string for exposure 2  \\
EXPID03 		& `01r-20130604001348-4-81283693' 	&	ID string for exposure 3  \\
EXPID04 		& `01r-20130604005109-5-81283731' 	&	ID string for exposure 4  \\
NEXP    		& 2 							&	Number of valid exposures  		\\
NEXP$\_$B  	& 2 							&	Number of valid blue exposures       \\
NEXP$\_$R  	& 2 							&	Number of valid red exposures        \\
EXPT$\_$B  	& 3600.00 						&	[s] Blue exposure duration time  	\\
EXPT$\_$R  	& 3600.00 						&	[s] Red exposure duration time		\\
EXPTIME 		& 3600.00 						&	[s] Minimum of exposure time for all cameras	\\
BESTEXP 		& 81283693 					&	MJM of the best exposure					\\
SCAMEAN 		& 2.11 						&	[ADU] Mean level of scatter light			\\
EXTRACT 		& `aperture'           			&	Extraction method						\\
SFLATTEN		& T 							&	Super flat has been applied				\\
PCASKYSB		& T 							&	PCA sky-subtraction has been applied			\\
NSKIES  		& 22 							&	Sky fiber number						\\
SKYCHI2 		& 1.91072039307 				&	${Mean chi^2 of sky-subtraction}$				\\
SCHI2MIN		& 1.51493247984 				&	${Minimum chi^2 of sky-subtraction}$			\\
SCHI2MAX		& 2.31929337901 				&    ${Maximum chi^2 of sky-subtraction}$			\\

\hline
\end{tabular}
\begin{flushleft}
{\sc Notes:}\\
\textbf{LAMPLIST} The value field shall contain a character string giving the file name of arc lamp emission line list, which is used in the process of wavelength calibration.\\
\textbf{SKYLIST}  The value field shall contain a character string giving the file name of sky emission line list, which is used in the process of sky subtraction.\\
\textbf{EXPID01, EXPID02, EXPID03, EXPID04} The value field of these four keywords contain four character strings giving the ID strings of first, second, third and fourth exposures respectively, which are analogue to the file names of LAMOST raw data of four exposures. It should be noted that, if the exposure times of a target are n, there should be n associated keywords here, they are EXPID01, EXPID02, ....EXPIDn respectively.\\
\textbf{NEXP, NEXP$\_$B, NEXP$\_$R} The value field of these three keywords shall contain three non-negative integers, which provide numbers of exposures, and numbers of valid blue and red exposures respectively.\\
\textbf{EXPT$\_$B, EXPT$\_$R} The value fields of these two keywords shall contain two real floating-point numbers, which give exposure duration times of blue and red CCD.\\
\textbf{EXPTIME} The value field shall contain a real floating-point, which gives the minimum of blue and red total exposures times.\\
\textbf{BESTEXP} The value field shall contain a integer, which gives the MJM of a exposure with maximum signal and noise ratio in n time exposures.\\
\textbf{SCAMEAN} The value field shall contain a real floating-point giving the mean level of scatter light, which is the average flux of regions where there is no fiber and is at the left and right edge of a two dimension spectra image.\\
\textbf{EXTRACT} The value field shall contain a character string, which indicates the method of spectrum extraction. In LAMOST spectra reduction pipeline, only the aperture method is applied to spectra extraction.\\
\textbf{SFLATTEN} The value of this keyword shall be Boolean, which represents whether or not use the super flat. In LAMOST spectra reduction pipeline, super flat is used to make the fiber-to-fiber relative efficiency around 1, and it can be estimated through spline fitting the flat flux of all fibers in a spectrograph.\\ 
\textbf{PCASKYSB} The value of this keyword shall be Boolean, which represents whether or not use the PCA method to subtract sky light. In LAMOST spectra reduction pipeline, the PCA method is used to subtract sky light at the wavelength range larger than 7200 angstrom.\\
\textbf{NSKIES} The value field shall contain a integer, which shows the number of sky fiber in a spectrograph.\\ 
\textbf{SKYCHI2} The value field shall contain a real floating-point, which gives the mean chi square of sky-subtraction. In the process of LAMOST spectra reduction, super sky is obtained by spline fitting m sky spectra. And thus, the chi square between the super sky and each sky spectra in an exposure, and the average chi square of m sky spectra can also be able to obtain. Assuming n times exposures, there will be 2n average chi square because of n blue spectra and n red spectra, and this keyword will be evaluated by calculating the mean value of these 2n average chi squares.\\
\textbf{SCHI2MIN} The value field shall contain a real floating-point, which gives the minimum chi square of sky-subtraction. As mentioned above, there will be 2n average chi squares assuming n time exposures, and this keyword will be the maximum of these chi squares.\\
\textbf{SCHI2MAX} The value field shall contain a real floating-point, which gives the maximum chi square of sky-subtraction. As mentioned above, there will be 2n average chi squares assuming n time exposures, this keyword will be the minimum of these chi squares.\\  
\end{flushleft}
\end{center}
\end{table}

\setcounter{table}{6}

%%%%%%%%%%%%%%%%%%%%%%% Part B of table 9

\begin{table}
\begin{center}
\caption{{Data Reduction Parameters Keywords(2)}\label{table0}}
\begin{tabular}[]{lllll}
\hline
Keywords 		&	Value						& 	Comment				\\
\hline
NSTD    		& 2 							& 	Number of (good) standard stars			\\
FSTAR   		& `149-182 '           			&	Fiber ID of flux standard stars			\\
FCBY    		& `catalog '           			&	Standard stars origin (auto, manual or catalog) \\
HELIO   		& T 							&	Heliocentric correction					\\
HELIO$\_$RV	& -3.99846411542 				&	[km/s] Heliocentric correction				\\
VACUUM  		& T 							&	Wavelengths are in vacuum					\\
NWORDER 		& 2 							& 	Number of linear-log10 coefficients			\\
WFITTYPE		& `LOG-LINEAR'         			&	Linear-log10 dispersion					\\
COEFF0  		& 3.56820 						&	Central wavelength (log10) of first pixel 	\\
COEFF1  		& 0.000100000 					&	Log10 dispersion per pixel				\\
WAT0$\_$001	& `system=linear'      			&									\\
WAT1$\_$001	& wtype=linear label=Wavelength &				\\
			& units=Angstroms    			&				\\
CRVAL1  		& 3.56820 						& Central wavelength (log10) of first pixel		\\
CD1$\_$1   	& 0.000100000 					& Log10 dispersion per pixel				\\
CRPIX1  		& 1 							& Starting pixel (1-indexed)				\\
CTYPE1  		& `LINEAR'           				&									\\
DC-FLAG 		& 1 							& Log-linear flag						\\
\hline
\end{tabular}
\begin{flushleft}
{\sc Notes:}\\
\textbf{NSTD} The value field shall contain a non-negative integer, which shows the number of flux standard stars with good spectra quality.\\
\textbf{FSTAR} The value field shall contain a character string giving the fiber identity numbers of flux standard stars, which are separated by the symbol ‘-’.\\
\textbf{FCBY} The value field shall contain a character string giving the selection methods of flux standard stars, which include auto, manual and catalog. Auto represents the standard stars are selected by the LAMOST reduction pipeline, manual means they are picked out by experienced staffs, and catalog indicates the standard stars are provided by the input catalog.\\ 
\textbf{HELIO} The value of this keyword shall be Boolean, which represents whether or not to perform the heliocentric correction. \\
\textbf{HELIO$\_$RV} The value field shall contain a real floating-point, which gives the radial velocity used to carry out the heliocentric correction.\\
\textbf{VACUUM} The value of this keyword shall be Boolean, which represents whether or not the LAMOST spectra is converted to vacuum wavelength.\\ 
\textbf{NWORDER} The value of this keyword shall contain a integer, which gives number of linear-log10 coefficients.\\
\textbf{WFITTYPE} The value field shall contain a character string giving linear-log10 dispersion.\\
\textbf{COEFF0} The value field shall contain a real floating-point number, which provides central wavelength (log10) of first pixel.\\
\textbf{COEFF1} The value field shall contain a real floating-point number giving log10 dispersion per pixel.\\
\textbf{WAT0$\_$001} The value field contain a character string.\\
\textbf{WAT1$\_$001} The value field contain a character string. \\
\textbf{CRVAL1} The value field shall contain a real floating-point number, which gives the coordinate value of the reference pixel provided by the CRPIX1 keyword[8].\\
\textbf{CD1$\_$1} The value field shall contain a real floating-point giving the dispersion of per pixel.\\ 
\textbf{CRPIX1} The value of this keyword shall contain a integer, which sets the reference pixel location on pixel axis[8].\\
\textbf{CTYPE1} The value field shall contain a character string, which will have the value ‘LINEAR’ to define the wavelength axes to be linear[9].\\
\textbf{DC-FLAG} The value of this keyword shall be Boolean, a value of 0 defines a linear sampling of the dispersion and a value of 1 defines a logarithmic sampling of the dispersion[9].\\
\end{flushleft}
\end{center}
\end{table}

%%%%%%%%%%%%%%%%%%%%%%%%%table10

\begin{table}
\begin{center}
\caption{{Spectra Analysis Results Keyword}\label{table10}}
\begin{tabular}[]{lllll}
\hline
Keywords 		&&	Value			&& 	Comment				\\
\hline
VERSPIPE		&&	 `v2.6.4  '     &&	Version of Pipeline		\\
CLASS   		&&	 `STAR    '    	&& 	Class of object		\\
SUBCLASS		&&	 `F5      '     &&	Subclass of object		\\
Z       		&&   -0.000314280000000 && Redshift of object		\\
Z$\_$ERR   	&&	 `        '     && 	Redshift error of object	\\
SN$\_$U   		&&    1.26 		&&	SNR of u filter		\\
SN$\_$G    	&&    4.68 		&&	SNR of g filter		\\
SN$\_$R    	&&    9.62 		&& 	SNR of r filter		\\
SN$\_$I    	&&    11.24 		&&   SNR of i filter		\\
SN$\_$Z    	&&    7.44 		&&   SNR of z filter		\\
\hline
\end{tabular}
\begin{flushleft}
{\sc Notes:}\\
\textbf{VERSPIPE} The value field shall contain a character string constant, which provides the version of LAMOST pipelines and are used to spectra processing and analysis. In this data release, the value of VERSPIPE is `v2.6.4’. It should be noted that, `v2.6’ is the version of spectra reduction pipeline, `v4’ is the version of spectra analysis pipeline, and `v2.6.4’ combines these two versions together.\\
\textbf{CLASS} The value field shall contain a character string providing the classification result determined by the LAMOST spectra analysis pipeline, which includes `STAR’, `GALAXY’, `QSO’ or `Unknown’.\\
\textbf{SUBCLASS} The value field shall contain a character string, which gives a sub-classification results for stars. This keyword provides a more detailed spectra type for late A, F, G, K and M, and spectra and photometric type for A type stars. For galaxies, quasars or unknown type objects, this keyword is set to `Non’ now. \\
\textbf{Z} The value field shall contain a real floating-point number providing redshift for a target, which is determined mainly by the LAMOST spectra analysis pipeline. For the case that redshift is unable to calculate by the pipeline, it will be manually determined through measuring the shifts of some spectral line centers. If the quality of a spectrum is poor, or it is classified as `Unknown’, its redshift is artificially set to -9999.0. \\
\textbf{Z$\_$ERR} The value field shall contain a real floating-point number, which gives redshift error of a target. At present, the values of Z$\_$ERR keywords are not yet published, and we will release redshift errors at the next data release in September of this year.\\
\textbf{SN$\_$U, SN$\_$G, SN$\_$R, SN$\_$I and SN$\_$Z} Keyword  The value fields of these five keywords shall contain five real floating-point numbers, which give the signal and noise ratio (SNR) of u, g, r, i and z bands. Using the center wavelength and band width, we can obtain the wavelength range of each SDSS band, and then the SNR in each band is the median value at each pixel in this band.\\
\end{flushleft}
\end{center}
\end{table}

%%%%%%%%%%%%%%%%%%%%%%%%%table11
\begin{table}
\begin{center}
\caption{{The significance of six bits of `Andmask' and `Ormask'}\label{table11}}
\begin{tabular}[The significance of six bits of `Andmask' and `Ormask']{lll}
\hline
Bit		&	Keyword	&	Comments\\
\hline
1	&	BADCCD		&	bad pixel on CCD			\\
2	&	BADPROFILE		&	bad profile in extraction		\\
3	&	NOSKY	no 		&	sky information at this wavelength\\
4	&	BRIGHTSKY		&	sky level too high			\\
5	&	BADCENTER		&	fiber trace out of the CCD		\\
6	&	NODATA		&	no good data				\\
\hline
\end{tabular}
\begin{flushleft}
{\sc Notes:}\\
1. The `andmask' information is a decimal integer determined by six-bit binary number, which represent six situations respectively as listed in this table. The associated bit of `andmask' will be set to 1, if the case always appears in each exposure. Like `andmask', `ormask' information is also a decimal integer determined by six-bit binary number. The difference is that each bit of `ormask' will be set to 1 if the related case happens in any exposure.\\
2. For a spectrum, if you want to check in which case has happened during reduction process, you should convert the decimal `Andmask’ and `Ormask’to six-bit binary number, and a case must has happened in all exposures(at least one exposure) if associated bit is 1 in binary `Andmask’(`Ormask').\\
\end{flushleft}
\end{center}
\end{table}

%%%%%%%%%%%%%%%%%%%%%%%%table12
\begin{table}
\begin{center}
\caption{{LAMOST general catalog}\label{table12}}
\begin{tabular}[LAMOST general catalog]{lll}
\hline
Field (unit)	&	Type	&	Comment\\
\hline
designation	&	varchar		&	Target Designation		\\
specObjId		&	long integer	&	Unique Spectra ID		\\
obsdate		&	char			&	Target Observation Date	\\
mjd			&	char			&	Local Modified Julian Day	\\
planid		&	char			&	Plan ID				\\
spid			&	integer		&	Spectrograph ID			\\
fiberid		&	integer		&	Fiber ID				\\
objra (degree)	&	float			&	Right Ascension			\\
objdec (degree) &	float			&	Declination			\\
dataversion	&	char			&	Data Version			\\
snru			&	float			&	Signal Noise Ratio of u filter\\
snrg			&	float			&	Signal Noise Ratio of g filter\\
snrr			&	float			&	Signal Noise Ratio of r filter\\
snri			&	float			&	Signal Noise Ratio of i filter\\
snrz			&	float			&	Signal Noise Ratio of z filter\\
objtype		&	varchar		&	Object Type			\\
class			&	varchar		&	Spectra Type			\\
subclass		&	varchar		&	Stellar Sub-Class		\\
magtype		&	varchar		&	Target Magnitude Type		\\
mag1 			&	(mag)	float		&	Associated Magnitude 1	\\
mag2 			&	(mag)	float		&	Associated Magnitude 2	\\
mag3 			&	(mag)	float		&	Associated Magnitude 3	\\
mag4 			&	(mag)	float		&	Associated Magnitude 4	\\
mag5 			&	(mag)	float		&	Associated Magnitude 5	\\
mag6 			&	(mag)	float		&	Associated Magnitude 6	\\
mag7 			&	(mag)	float		&	Associated Magnitude 7	\\
tsource		&	varchar		&	Organization or person who submit input catalog\\
tfrom			&	varchar		&	Input catalog submitted by an organization or a person\\
fibertype		&	varchar		&	Fiber Type of target[Obj, Sky, F-std, Unused, PosErr, Dead]\\
z			&	float			&	Redshift				\\
rv 			&	(km/s)float	&	Heliocentric Radial Velocity\\
\hline
\end{tabular}
\end{center}
\end{table}

%%%%%%%%%%%%%%%%%%%%%%%%table13
\begin{table}
\begin{center}
\caption{{A type stars catalog}\label{table13}}
\begin{tabular}[A type stars catalog]{lll}
\hline
Field (unit)	&	Type	&	Comment\\
\hline
designation	&	varchar		&	Target Designation		\\
specObjId		&	long integer	&	Unique Spectra ID		\\
obsdate		&	char			&	Target Observation Date	\\
mjd			&	char			&	Local Modified Julian Day	\\
planid		&	char			&	Plan ID				\\
spid			&	integer		&	Spectrograph ID			\\
fiberid		&	integer		&	Fiber ID				\\
objra (degree)	&	float			&	Right Ascension			\\
objdec (degree)	&	float			&	Declination			\\
dataversion	&	char			&	Data Version			\\
snru			&	float			&	Signal Noise Ratio of u filter\\
snrg			&	float			&	Signal Noise Ratio of g filter\\
snrr			&	float			&	Signal Noise Ratio of r filter\\
snri			&	float			&	Signal Noise Ratio of i filter\\
snrz			&	float			&	Signal Noise Ratio of z filter\\
objtype		&	varchar		&	Object Type			\\
class			&	varchar		&	Stellar Class			\\
subclass		&	varchar		&	Stellar Sub-Class		\\
magtype		&	varchar		&	Target Magnitude Type		\\
mag1 (mag)		&	float			&	Associated Magnitude 1	\\
mag2 (mag)		&	float			&	Associated Magnitude 2	\\
mag3 (mag)		&	float			&	Associated Magnitude 3	\\
mag4 (mag)		&	float			&	Associated Magnitude 4	\\
mag5 (mag)		&	float			&	Associated Magnitude 5	\\
mag6 (mag)		&	float			&	Associated Magnitude 6	\\
mag7 (mag)		&	float			&	Associated Magnitude 7	\\
tsource		&	varchar		&	Organization or person who submit input catalog\\
tfrom			&	varchar		&	Input catalog submitted by an organization or a person\\
fibertype		&	varchar		&	Fiber Type of target[Obj, Sky, F-std, Unused, PosErr, Dead]\\
rv (km/s)		&	float			&	Heliocentric Radial Velocity\\
Halpha$\_$Indice (angstrom)	& float	&	Line Indice of Halpha line   \\
Hbeta$\_$Indice (angstrom)	& float	&	Line Indice of Hbeta line		\\
Hgama$\_$Indice (angstrom)	& float	&	Line Indice of Hgama line		\\
Hdelta$\_$Indice (angstrom)	& float	&	Line Indice of Hdelta line		\\
Halpha$\_$D0.2 (angstrom)	& float	&	Width at 20$\%$ below the local continuum of Halpha line\\
Hbeta$\_$D0.2 (angstrom)		& float	&	Width at 20$\%$ below the local continuum of Hbeta line\\
Hgama$\_$D0.2 (angstrom)		& float	&	Width at 20$\%$ below the local continuum of Hgama line\\
Hdelta$\_$D0.2 (angstrom)	& float	&	Width at 20$\%$ below the local continuum of Hdelta line\\
\hline
\end{tabular}
\end{center}
\end{table}

%%%%%%%%%%%%%%%%%%%%%%%%table14
\begin{table}
\begin{center}
\caption{{A, F, G and K type stars catalog}\label{table14}}
\begin{tabular}[A, F, G and K type stars catalog]{lll}
\hline
Field (unit)	&	Type		&	Comment\\
\hline
designation	&	varchar		&	Target Designation								\\
specObjId		&	long integer	&	Unique Spectra ID								\\
obsdate		&	char			&	Target Observation Date							\\
mjd			&	char			&	Local Modified Julian Day							\\
planid		&	char			&	Plan ID										\\
spid			&	integer		&	Spectrograph ID									\\
fiberid		&	integer		&	Fiber ID										\\
objra (degree)	&	float			&	Right Ascension									\\
objdec (degree)	&	float			&	Declination									\\
dataversion	&	char			&	Data Version									\\
snru			&	float			&	Signal Noise Ratio of u filter						\\
snrg			&	float			&	Signal Noise Ratio of g filter						\\
snrr			&	float			&	Signal Noise Ratio of r filter						\\
snri			&	float			&	Signal Noise Ratio of i filter						\\
snrz			&	float			&	Signal Noise Ratio of z filter						\\
objtype		&	varchar		&	Object Type									\\
class			&	varchar		&	Stellar Class									\\
subclass		&	varchar		&	Stellar Sub-Class								\\
magtype		&	varchar		&	Target Magnitude Type								\\
mag1 (mag)		&	float			&	Associated Magnitude 1							\\
mag2 (mag)		&	float			&	Associated Magnitude 2							\\
mag3 (mag)		&	float			&	Associated Magnitude 3							\\
mag4 (mag)		&	float			&	Associated Magnitude 4							\\
mag5 (mag)		&	float			&	Associated Magnitude 5							\\
mag6 (mag)		&	float			&	Associated Magnitude 6							\\
mag7 (mag)		&	float			&	Associated Magnitude 7							\\
tsource		&	varchar		&	Organization or person who submit input catalog			\\
tfrom			&	varchar		&	Input catalog submitted by an organization or a person\\
fibertype		&	varchar		&	Fiber Type of target[Obj, Sky, F-std, Unused, PosErr, Dead]\\
rv (km/s)		&	float			&	Heliocentric Radial Velocity						\\
teff (K)		&	float			&	Effective Temperature								\\
log (dex)		&	float			&	Surface Gravity									\\
feh (dex)		&	float			&	Metallicity									\\
\hline
\end{tabular}
\end{center}
\end{table}

%%%%%%%%%%%%%%%%%%%%%%%%table15
\begin{table}
\begin{center}
\caption[]{{M-type star catalog}\label{table15}}
\begin{tabular}[M catalog]{lll}
\hline
Field (unit)	&	Type	&	Comment\\
\hline
designation	&	varchar		&	Target Designation	\\
specObjId		&	long integer	&	Unique Spectra ID	\\
obsdate		&	float			&	Target Observation Date\\
mjd			&	char			&	Local Modified Julian Day\\
planid		&	char			&	Plan ID			\\
spid			&	integer		&	Spectrograph ID		\\
fiberid		&	integer		&	Fiber ID			\\
objra (degree)	&	float			&	Right Ascension		\\
objdec (degree)	&	float			&	Declination		\\
dataversion	&	char			&	Data Version		\\
snru			&	float			&	Signal Noise Ratio of u filter\\
snrg			&	float			&	Signal Noise Ratio of g filter\\
snrr			&	float			&	Signal Noise Ratio of r filter\\
snri			&	float			&	Signal Noise Ratio of i filter\\
snrz			&	float			&	Signal Noise Ratio of z filter\\
objtype		&	varchar		&	Object Type				\\
class			&	varchar		&	Stellar Class				\\
subclass		&	varchar		&	Stellar Sub-Class			\\
magtype		&	varchar		&	Target Magnitude Type			\\
mag1 (mag)		&	float			&	Associated Magnitude 1		\\
mag2 (mag)		&	float			&	Associated Magnitude 2		\\
mag3 (mag)		&	float			&	Associated Magnitude 3		\\
mag4 (mag)		&	float			&	Associated Magnitude 4		\\
mag5 (mag)		&	float			&	Associated Magnitude 5		\\
mag6 (mag)		&	float			&	Associated Magnitude 6		\\
mag7 (mag)		&	float			&	Associated Magnitude 7		\\
tsource		&	varchar		&	Organization or person who submit input catalog	\\
tfrom			&	varchar		&	Input catalog submitted by an organization or a person \\
fibertype		&	varchar		&	Fiber Type of target[Obj, Sky, F-std, Unused, PosErr, Dead]\\
rv (km/s)		&	float			&	Heliocentric Radial Velocity	\\
magact		&	integer		&	Magnetic Activity			\\
\hline
\end{tabular}
\end{center}
\end{table}

\label{lastpage}

\end{document}